\documentclass[aps,pre,twocolumn,showpacs,superscriptaddress,floatfix]{revtex4}
\usepackage{graphics,amssymb,amsmath,epsfig,ulem} 

\begin{document} 

\title{Glucose Metabolism and Oscillatory Behavior of Pancreatic Islets} 

\author{H. Kang}
\affiliation{Department of Physics and Center for Theoretical Physics,
Seoul National University, Seoul 151-747, Korea}
\author{J. Jo}
\affiliation{Department of Physics and Center for Theoretical Physics,
Seoul National University, Seoul 151-747, Korea}
\author{H.J. Kim}
\affiliation{Department of Physics and Center for Theoretical Physics,
Seoul National University, Seoul 151-747, Korea}
\author{M.Y. Choi}
\affiliation{Department of Physics and Center for Theoretical Physics,
Seoul National University, Seoul 151-747, Korea}
\affiliation{Korea Institute for Advanced Study, Seoul 130-722, Korea}

\author{S.W. Rhee}
\affiliation{Department of Physiology, University of Arkansas, Little Rock, AR 72205, U.S.A.}

\author{D.S. Koh}
\affiliation{Department of Physics, Pohang University of Science and Technology,
Pohang 790-784, Korea}
\affiliation{Department of Physiology, University of Washington, Seattle, WA 98195, U.S.A.}

\begin{abstract}
A variety of oscillations are observed in pancreatic islets.
We establish a model, incorporating two oscillatory systems of different time scales:
One is the well-known bursting model in pancreatic $\beta$-cells and the other is
the glucose-insulin feedback model which considers direct and indirect feedback
of secreted insulin.
These two are coupled to interact with each other in the combined model, and
two basic assumptions are made on the basis of biological observations:
The conductance $g_{K(ATP)}$ for the ATP-dependent potassium current is a decreasing
function of the glucose concentration whereas the insulin secretion rate is given
by a function of the intracellular calcium concentration.
Obtained via extensive numerical simulations are complex oscillations including
clusters of bursts, slow and fast calcium oscillations, and so on.
We also consider how the intracellular glucose concentration depends upon the extracellular
glucose concentration, and examine the inhibitory effects of insulin.
%
\end{abstract}


\maketitle

\section{Introduction}
Insulin secretion from pancreatic $\beta$ cells is critical for glucose homeostasis in blood.
Perturbations from the basal glucose concentration induce various oscillations with different
periods in pancreatic $\beta$ cells: fast oscillations with periods of several
seconds~\cite{Sherman}, slow oscillations with periods of several minutes~\cite{Maki},
and ultradian oscillations with periods of a few hours~\cite{Sturis}.
Each type of oscillations has been studied separately, within the respective model
according to each time scale. For example, slow oscillations can be caused by a feedback
mechanism having slow reaction time~\cite{Maki}.
Independently of this, a microscopic model for the bursting mechanism of the membrane
potential in a $\beta$ cell was devised and investigated~\cite{Sherman}
while the bursting activity was also demonstrated in a simple two-dimensional
map~\cite{de Vries}.
In particular, the microscopic bursting model, which was constructed in a similar manner to
the Hodgkin-Huxley model for nerve excitation~\cite{Hodgkin}, was improved by the
introduction of the potassium ion K$^+$ conductance depending dynamically on the ATP
concentration~\cite{Keizer}.
However, the correlations between the bursting mechanism and insulin secretion have not
been addressed in an appropriate way and there still lacks a model describing the complex
phenomena such as clusters of bursting action potentials~\cite{Lefebvre}, fast and slow
Ca$^{2+}$ oscillations~\cite{Liu}, which arise from combined behaviors of different time
scales. The purpose of this study is to propose a combined model, explaining both fast and
slow oscillations of action potentials.

\begin{figure}
\vspace{1cm}
\epsfig{width=7.16cm,file=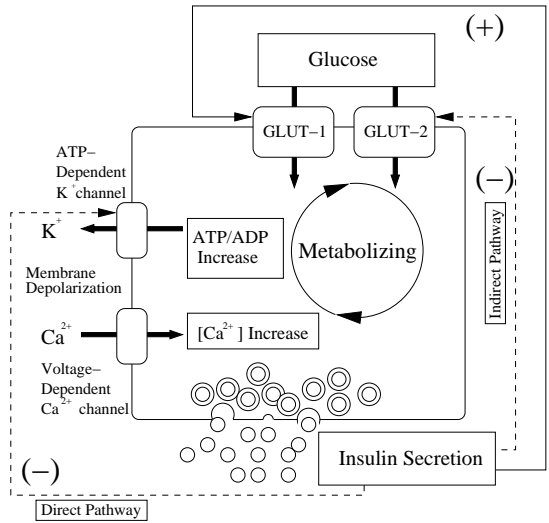} \\
\caption{Mechanism of insulin secretion.
Activation and inhibition of GLUT-1 and GLUT-2 transporters by secreted insulin are
represented by the solid ($+$) and dashed ($-$) arrows.
Thick arrows describe physical transport of materials (glucose and ions).
}
\label{metabolism}
\end{figure}

On the other hand, recent experimental investigations in vivo and in vitro
about the oscillations of insulin secretion have revealed a more complex picture
on the molecular basis. Stimulation of insulin secretion by glucose involves a rise
in the cytoplasmic concentration of calcium ions (Ca$^{2+}$) in $\beta$ cells.
This rise essentially results from the following sequence of events,
as sketched in Fig.~\ref{metabolism}:
Glucose transported by GLUT-1 and GLUT-2 transporters in a $\beta$ cell raises
the ratio of adenosine triphosphate (ATP) to adenosine diphosphate (ADP),
which promotes closure of ATP-sensitive K$^+$ channels.
This generates membrane depolarization, urging voltage-dependent
calcium channels to open. The subsequent increase in free cytosolic Ca$^{2+}$
then stimulates insulin secretion~\cite{Ashcroft}.
Note also that there are several experimental evidences for insulin to inhibit
bursting of $\beta$ cells~\cite{Persaud,Khan};
in particular, there is a report that insulin activates ATP-sensitive K$^+$ channels
in $\beta$ cells~\cite{Nunemaker}.  Based on these experiments, we assume two
possible inhibitory pathways of insulin:
indirect effects on gating of K(ATP) channels by inhibiting glucose transport
through GLUT-2 transporters by insulin and direct effects of insulin to activate the channel.
Both negative feedback processes of insulin to activate the channels may result in
oscillatory behavior.
Oscillations of the membrane potential~\cite{Andres} drive oscillations
of Ca$^{2+}$~\cite{Dryselius},
leading in turn to oscillations of insulin secretion~\cite{Hellman,Gilon}.
In fact, the measured portal-vein insulin secretion rate shows periodic behavior of
insulin secretion with successive secretion and rest.
The frequency of such pulsatile insulin secretion in vivo and that in vitro
in an isolated perfused pancreas have been reported to vary from four to fifteen minutes
and from six to ten minutes, respectively~\cite{Chou,Porksen}.

These experimental results make it desirable to establish a model
that can describe the whole feedback process on the microscopic level,
with the aforementioned macroscopic features incorporated.
As an attempt toward such a goal, we propose a model, connecting
the macroscopic description of glucose regulation~\cite{Maki}
and the microscopic mechanism of bursting behavior of $\beta$ cells~\cite{Sherman}
via the prescription based on experimental observations:
The conductance $g_{K(ATP)}$ for the K(ATP) current, i.e., the K$^+$ current through the
ATP-dependent K$^+$ channel, is taken to be a decreasing function of the intracellular
glucose concentration whereas it is assumed that the insulin secretion rate
depends on the intracellular Ca$^{2+}$ concentration.
For the macroscopic description, we consider two possible insulin induced inhibitory pathways
affecting the conductance $g_{K(ATP)}$.
In addition, electrical coupling through gap junctions between $\beta$ cells
in an islet is assumed.
Obtained via extensive numerical simulations are a variety of oscillatory
behavior of the bursting electrical activity, calcium concentration,
and insulin secretion, which are in good agreement with those observed experimentally.

There are four sections in this paper: Section II introduces the model system,
described by coupled differential equations, together with the appropriate parameters.
The coupled equations are integrated numerically and the results are presented in Sec. III.
Finally, a detailed discussion, together with a brief summary, is given in Sec. IV.

\section{Model System}

To describe the whole feedback process, we incorporate the bursting mechanism of $\beta$
cells given by Sherman~\cite{Sherman} and the description of glucose-insulin oscillations
by Maki and Keizer~\cite{Maki,Keener}.
We begin with the dynamic equation for the time-dependent
behavior of the membrane potential $V_i$ of the $i$th $\beta$ cell~\cite{Sherman}:
\begin{eqnarray}
\label{V}
\tau \frac{dV_i}{dt} = &-& {I_i}_{Ca} - {I_i}_K - {I_i}_S - {I_i}_{K(ATP)} \nonumber \\
&-& {\sum_j}^{\prime} g_c (V_i - V_j),
\end{eqnarray}
which describes current balance with the relaxation time $\tau$.
The right-hand side of Eq.~(\ref{V}) includes contributions from various
current channels, which are all given in units of voltage 
(i.e., with the dimensionless conductance as below): 
calcium current $I_{Ca}$, potassium current $I_K$, generic slow 
current $I_S$, and background current $I_{K(ATP)}$ 
through the ATP-dependent potassium channel. 
The last term, where the prime in the summation 
stands for the restriction that $j$ be a nearest neighbor of $i$, 
represents the electrical coupling via gap junctions of 
(dimensionless) conductance $g_c$ 
between nearest neighboring cells in an islet. 

The first three types of current depend on the membrane potential $V_i$ 
and on the gating variables $N_i$ or $S_i$, and are given by 
\begin{eqnarray*} 
{I_i}_{Ca} & = & g_{Ca} M_{\infty}(V_i - V_{Ca}) \\ 
{I_i}_K & = & g_K N_i (V_i - V_K) \\ 
{I_i}_S & = & g_S S_i (V_i - V_K) 
\end{eqnarray*} 
with constant (dimensionless) conductances $g_{Ca}$, $g_K$, and $g_S$ 
and the reversal potentials $V_{Ca}$ and $V_K$ for Ca$^{2+}$ and K$^+$ ions, respectively. 
The gating variables $N_i$ and $S_i$ are governed by the equations 
\begin{eqnarray} 
\tau \frac{dN_i}{dt} & =& \lambda ( N_{\infty} - N_i ) \label{N}\\ [1mm] 
\tau_{S} \frac{dS_i}{dt} & = & S_{\infty} - S_i \label{S} 
\end{eqnarray} 
where the activation values $N_{\infty}$ and $S_{\infty}$ as well as $M_{\infty}$ 
in general depend on the membrane potential. 
Thus the calcium current has been assumed to respond instantaneously 
to a change in the membrane potential via the voltage-dependent 
activation $M_{\infty}$ while the potassium current is governed by
the gating variable $N_i$ via Eq.~(\ref {N}).
These two currents are responsible for generating the action potential
during the active phase of bursting.
The generic slow current, which may corresponds to
the inhibitory potassium current, is gated by the slow variable $S_i$
via Eq.~(\ref {S}) and responsible for switching between the active
and silent phases.
We use Boltzmann-type expressions for the equilibrium values $M_{\infty}$,
$N_{\infty}$, and $S_{\infty}$ of the voltage-dependent activation:
$$
X_{\infty} = \frac{1}{1 + \exp{[(V_X - V_i) / \theta_X]}}
$$
with appropriate constants $V_X$ and $\theta_X$,
where $X$ denotes $M$, $N$, or $S$.

On the other hand, $I_{K(ATP)}$, which is responsible for setting the plateau fraction, i.e.,
the ratio of the active phase duration to the burst period, satisfies
$$
{I_i}_{K(ATP)} = g_{K(ATP)} p_i (V_i - V_K),
$$
where the conductance $g_{K(ATP)}$ depends on the glucose concentration
and $p_i$ is the opening probability of the K(ATP) channel.
In general, an increase in the intracellular glucose concentration due to
injection of glucose raises the ATP/ADP ratio after being metabolized;
this in turn reduces $g_{K(ATP)}$ and induces depolarization of the membrane
potential~\cite{Gopel}.
Considering the time delay in this process, i.e., between the increase of the intracellular
glucose concentration $G_{in}$ and the reduction of the conductance $g_{K(ATP)}$,
we assume that $g_{K(ATP)}$ at time $t$ is a decreasing function of $G_{in}$ at time
$t{-}t_d$, where the delay time $t_d$ measures the time needed for metabolizing
glucose transported into the $\beta$ cell.
To be specific, we take the Hill equation~\cite{Hill}:
$$
g_{K(ATP)}(t) = \frac{ g_1 - g_2 }{[G_{in}(t-t_d)/G_K]^b f_D(J_i) +1} + g_2,
$$
with the appropriate Hill coefficient $b$~\cite{index}.
Note the sharp contrast to the constant $g_{K(ATP)}$ in existing studies on
the bursting mechanism~\cite{Sherman} or to the case that $g_{K(ATP)}$ is taken to
be a function of ATP~\cite{Keizer}.
Here $g_1$ and $g_2$ represent the maximum and minimum values of $g_{K(ATP)}$,
respectively, whereas $G_K$ is the value of $G_{in}$ for which $g_{K(ATP)}$ reduces to
its medium value $(g_1+g_2)/2$.
The function $f_D (J_i)$ describes possible direct effects of insulin on $g_{K(ATP)}$.
Namely, while we have simply $f_D = 1$ in the absence of such direct effects,
in their presence we take $f_D (J_i)$ to be a decreasing function $f(J_i)$
of the inhibition variable $J_i$:
\begin{equation*}
f(J_i) = 1 - \frac{1+{J_0}^m}{1+(J_0 /J_i)^m}
\end{equation*}
with the inhibition variable governed by
\begin{equation}
\label{J}
\tau_J \frac{dJ_i}{dt} = J_{\infty} - J_i .
\end{equation}
The (slow) relaxation time $\tau_J$ and the constant $J_0$ depend on the pathway,
whereas $J_{\infty}$ is taken to be an increasing function of 
the insulin concentration $H_i$: 
\begin{eqnarray*} 
J_{\infty} = \frac{H_i}{H_J + H_i} 
\end{eqnarray*} 
with appropriate constant $H_J$ (see Table I). 

Although there are many sources of noise in a biological system,
we here restrict ourselves only to the stochastic opening and closing of ion channels, 
in particular those of K(ATP) channels, and assume that 
the dynamics of the opening probability $p_i$ in the $i$th cell is described 
by the stochastic differential equation~\cite{Sherman,Fox} 
\begin{equation} 
\label{p} 
\frac{dp_i}{dt} = \frac{\gamma_1}{\tau_{p}} (1 - {p_i}) 
- \frac{\gamma_2}{\tau_{p}} p_i + \eta_{i}(t) 
\end{equation} 
with appropriate constants $\gamma_1$ and $\gamma_2$ and relaxation time $\tau_p$, 
where $\eta_i (t)$ is the Gaussian white noise with zero mean and variance 
$$ 
\langle \eta_{i}(t) \eta_{j}(t') \rangle = 
\frac{\gamma_1 (1 - {p_i}) + \gamma_2 p_i}{\tau_{p} n_{K(ATP)}} 
\delta_{ij} \delta (t - t^{\prime}). 
$$ 
Note that as the number $n_{K(ATP)}$ of K(ATP) channels per cell 
grows large ($n_{K(ATP)} \rightarrow \infty$), we have 
$\eta_i \rightarrow 0$ and obtain the usual deterministic equation for $p_i$. 

Modifying the model for a perifusion system~\cite{Maki}, 
we now obtain equations describing glucose regulation. 
In view of the perifusion, we consider conservation of insulin 
and write the equation for the insulin concentration $H_i$ around the $i$th cell 
in the form: 
\begin{equation} 
\label{I} 
\frac{dH_i}{dt} = \frac{R_s}{\Omega} - k_0 (H_i - H_0), 
\end{equation} 
where $R_s$ is the rate of insulin secretion, $k_0$ is the flow rate for each cell, 
and $H_0$ is the background concentration. 
The effective volume $\Omega$ of a $\beta$ cell is given by the volume of the islet 
divided by the number of $\beta$ cells in it. 
Experimental observations indicate that insulin secretion from $\beta$ cells
increases with the intracellular calcium concentration
and with the amount of insulin stored for rapid secretion~\cite{Rorsman}.
In particular, the secretion rate is known to depend on the Ca$^{2+}$
concentration quartically.
We thus assume a quartic function of the concentration $C_i$ of
the free intracellular Ca$^{2+}$:
$$
R_s = R_0 {\tilde{C}_i}^4 \frac{S_{R}}{S_{max}},
$$ 
where $R_0$ sets the scale, $\tilde{C}_i$ is the rescaled (dimensionless) calcium 
concentration defined to be $\tilde{C}_i \equiv (C_i - C_b)/C_0$ 
with the background value $C_b$ and the appropriate scale $C_0$, 
and $S_{R}$ is the storage amount of insulin preparing rapid secretion 
in the readily releasable vesicle pool (RRVP) of the $\beta$-cell, 
with the maximum value $S_{max}$. 
The latter changes according to the relation 
\begin{equation} 
\frac{dS_{R}}{dt} = R_r - R_s, 
\end{equation} 
where the refilling rate $R_r$ of insulin granules to the RRVP is proportional to the
remaining amount: 
$$ 
R_r = a_r (S_{max} - S_{R}) 
$$
with appropriate constant $a_r$. 
The Ca$^{2+}$ concentration $C_i$ in the $i$th cell 
is governed by the rate equation: 
\begin{equation} 
\label{C} 
\frac{dC_i}{dt} = - \epsilon(\alpha {I_i}_{Ca} - M_S - M_P - M_N + M_{leak}), 
\end{equation} 
where $\epsilon$ is the fraction of free Ca$^{2+}$ ions in the cytosolic compartments, 
$\alpha$ is the proportionality constant between the current flow and the 
concentration reduction rate. 
Clearance terms, $M_S$, $M_P$, and $M_N$ are operated by 
sarco-endoplasmic reticulum Ca$^{2+}$-ATPase (SERCA) pumps, plasma-membrane Ca$^{2+}$ ATPase, 
and plasma-membrane Na/Ca$^{2+}$ exchangers, respectively~\cite{Chen}, 
and given by 
\begin{eqnarray*} 
M_S &=& M_{S}^{max} \frac {1}{1+(K_S/C_i)^2} \\ 
M_P &=& M_{P}^{max} \frac {1}{1+(K_P/C_i)} \frac{[\mbox{H}^+]}{[\mbox{H}^+]+[\mbox{K}_a]} \\ 
M_N &=& k_{N} C_i 
\end{eqnarray*} 
with appropriate constants $M_{S}^{max}$, $M_{P}^{max}$, $k_{N}$, $K_S$, and $K_P$ 
as well as [H$^+$] and [K$_a$] adjusted to pH$ = 7.40$ and pK$_a = 7.86$ 
for activation of pumping by protons. 
The last term $M_{leak}$ represents the flux out of the endoplasmic reticulum (ER) 
and is described by
\begin{equation*} 
M_{leak} = \xi_{er}( C_i^{er} - C_i ), 
\end{equation*} 
where $\xi_{er}$ is the rate of Ca$^{2+}$ release from the ER 
and $C_i^{er}$ denotes the Ca$^{2+}$ concentration in the ER. 
The latter is governed by the rate equation 
\begin{equation} 
\frac{dC_i^{er}}{dt} = \epsilon_{er}\frac{\nu_{cyt}}{\nu_{er}}(M_S - M_{leak}), 
\end{equation} 
where $\epsilon_{er}$ is the fraction of free Ca$^{2+}$ ions in the ER, 
$\nu_{cyt}$ and $\nu_{er}$ are the volumes of the cytosolic and of the ER compartments, 
respectively~\cite{BertSher}. 

Finally, we consider the rate equation for the glucose concentration $G_{in}$ 
in the $i$th cell: 
\begin{equation} 
\label{Gbeta} 
\frac{dG_{in}}{dt} = r_1 + r_2 - r_m ,
\end{equation} 
where $r_{1, 2}$ are the rates of glucose uptake through GLUT-1 and GLUT-2 
transporters, respectively, and $r_m$ is the rate of glucose metabolism~\cite{index}. 
Here the difference between the glucose concentration in blood plasma and 
the injected glucose concentration $G_0$ has been disregarded for convenience, 
with the assumption that the former saturates rather quickly to the latter. 
We thus take the rates through GLUT-1 and GLUT-2 transporters to be 
simple increasing functions of the extracellular glucose concentration $G_0$ 
and choose an increasing function of the intracellular glucose concentration $G_{in}$ 
for the metabolism rate: 
\begin{eqnarray*} 
r_1 & = & \frac{c_1 (G_0-G_{in})}{(1+G_0/K_1)(K_1 + G_{in})} \frac{H_i}{H_r + H_i} \\ [1mm] 
r_2 & = & \frac{c_2 (G_0 -G_{in})f_I(J_i)}{(1+G_0/K_2)(K_2 + G_{in})} \\ [1mm]
r_m & = & \frac{c_m G_{in}}{K_m + G_{in}}. 
\end{eqnarray*} 
The constants $K_1$, $c_2$, $K_2$, $c_m$, and $K_m$ are determined from experimental results, 
whereas the constants $c_1$ and $H_r$ are chosen to yield appropriate intracellular glucose
concentrations in simulations. 
Note that $r_1$ has been taken to be an increasing function of $H_i$, 
which models the recruitment of GLUT-1 transporters by insulin 
and results in positive feedback. 
On the other hand, to take into account indirect effects of insulin on $g_{K(ATP)}$, 
$r_2$ has been chosen to depend on the function $f_I$ of the inhibition variable $J_i$, 
which accounts for inhibitory effects of insulin on its own release
via the inhibition of GLUT-2;
this assumption is made as a possible interpretation of the recent observation of
the inhibiting role of insulin~\cite{Persaud,Khan,Nunemaker}.
%
Accordingly, $r_2$ tends to decrease as the insulin concentration is raised.
Note in particular that unlike the form in the existing study~\cite{Maki},
$r_2$ in the above form always remains positive, thus avoiding the unphysiological
case of negative values.
To complete the model, we assume the two-step process in a $\beta$ cell,
which includes manifestation of the inhibitory signal from insulin receptors
in the presence of insulin (first step) and the inhibition of
GLUT-2 transporters by that signal
(second step);
this provides the slow negative feedback process, which may again be described by the
decreasing function $f(J_i)$ (for the direct effects before) with appropriate values of
the parameters.

The whole feedback model is thus described by the ten coupled equations
[Eqs. (\ref{V}) - (\ref{Gbeta})] for the ten variables,
together with the equations giving dependence on those variables.
It should be stressed that the bursting mechanism and the glucose
regulation are incorporated, with the glucose-dependent conductance
$g_{K(ATP)}$ and the calcium-dependent rate of insulin secretion
as well as the insulin-dependent rate of glucose uptake
playing the role of a bridge between the two.
The appropriate values of the parameters used in the whole model are
given in Table I.

\begin{table*}
\caption{Parameter values in the whole feedback model:
The parameters without asterisks or single asterisks are fixed or adjustable ones,
determined from experiments, respectively.
Those with double asterisks are free parameters, not determined from experiments.
}
\begin{ruledtabular}
\begin{tabular}{ccccc}
$g_{Ca} = 3.6$&$V_{Ca} = 25\,$mV&$V_M = - 20\,$mV
&${\theta}_M = 12\,$mV&$\tau = 20\,$ms
\\$g_{K} = 10$&$V_{K} = - 75\,$mV&$V_N = - 17\,$mV
&${\theta}_N = 5.6\,$mV&$\lambda = 0.8$
\\$^*g_{S} = 4$&$^*\gamma_1 = 1$&$^*V_S = - 22\,$mV
&$^*{\theta}_S = 8.0\,$mV&$^*{\tau}_S = 60\,$s
\\$\alpha = 1.3\times10^{-6}\,$M/Vs&$^*\gamma_2 = 1$
&$^*{\tau}_p = 0.30\,$s&$^*n_{K(ATP)} = 10^3$&$\Omega=1.5\times 10^{-15}\,$m$^3$
\\$\epsilon = 0.01$&$\epsilon_{er} = 0.01$&$^*\xi_{er} = 1.0\times 10^{-4}$
&$\nu_{cyt}= 10\,\mu$m$^3$&$\nu_{er}= 0.4\,\mu$m$^3$
\\$g_1 = 3.0$&$g_2 = 0.6$&$G_K = 2.8\,$mM&$b = 2.5$
&$^*S_{max} = 3.6\times 10^{-17}\,$mol
\\$^*M_S^{max} = 246\,$$\mu$M/s&$^*M_P^{max} = 126\,$$\mu$M/s
&$^*k_N = 84\,$s$^{-1}$&$K_S = 0.27\,\mu$M&$K_P = 0.50\,\mu$M
\\$c_m = 4.1\times 10^{-5}\,$M/s&$K_m = 7.8\,$mM&$^*R_0 = 1.2\times 10^{-19}\,$mol/s
&$^*a_r = 2.0\times 10^{-3}\,$s$^{-1}$&$[H^+] = 10^{-7.40}\,$M
\\$c_1 = 5.8\times 10^{-4}\,$M/s&$K_1 = 1.4\,$mM&$c_2 = 5.3\times 10^{-4}\,$M/s
&$K_2 = 17\,$mM&$[K_a] = 10^{-7.86}\,$M
\\$^{**}k_0 = 1.8\times 10^{-2}\,$s$^{-1}$&$^{**}H_r = 1.4\times 10^{-2}\,$mM
&$^{**}H_J = 1.4\times10^{-4}\,$mM&$^{**}t_d = 90\,$s&$^{**}m = 4.0\,$
\\$^{**}H_0 = 0\,$mM&$^*C_0 = 10^{-4}\,$mM&$^*C_b = 6.0\times 10^{-5}\,$mM&&
\end{tabular}
\end{ruledtabular}
\end{table*}

\section{Simulation Results}

We integrate numerically the ten equations, with the parameter values
given by Table I, by means of the fifth-order Runge-Kutta method. 
To keep track of the state of every single channel is computationally
expensive; it is sufficient to monitor only the total number of 
open and closed channels per cell. 
The time step $\Delta t$ in the numerical integration should be 
chosen sufficiently small so that the probability of 
two channel openings (or closings) during one time step is negligible, 
say, less than $0.1$~\cite{Sherman}: 
$$ 
\Delta t < 0.1\, {\rm{min}} 
\left[ \frac{\tau_p}{\gamma_1 n_{K(ATP)}}, \,
\frac{\tau_p}{\gamma_2 n_{K(ATP)}} \right]. 
$$ 
In order to control efficiently the error in the integration, 
we here use the adaptive step-size algorithm with 
$\Delta t_{max}=1 \,\rm{ms}$, keeping the above restriction on $\Delta t$. 
This controls automatically the maximum of the next step size 
to achieve the predetermined accuracy. 

\begin{figure} 
\epsfig{width=5cm,file=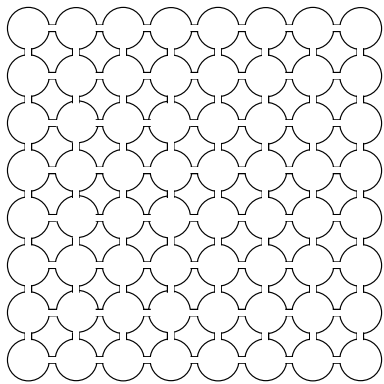} \\ 
\caption{Two-dimensional square lattice} 
\label{lattice} 
\end{figure} 

\begin{figure} 
\begin{tabular}{cccc} 
\epsfig{width=4.2cm,file=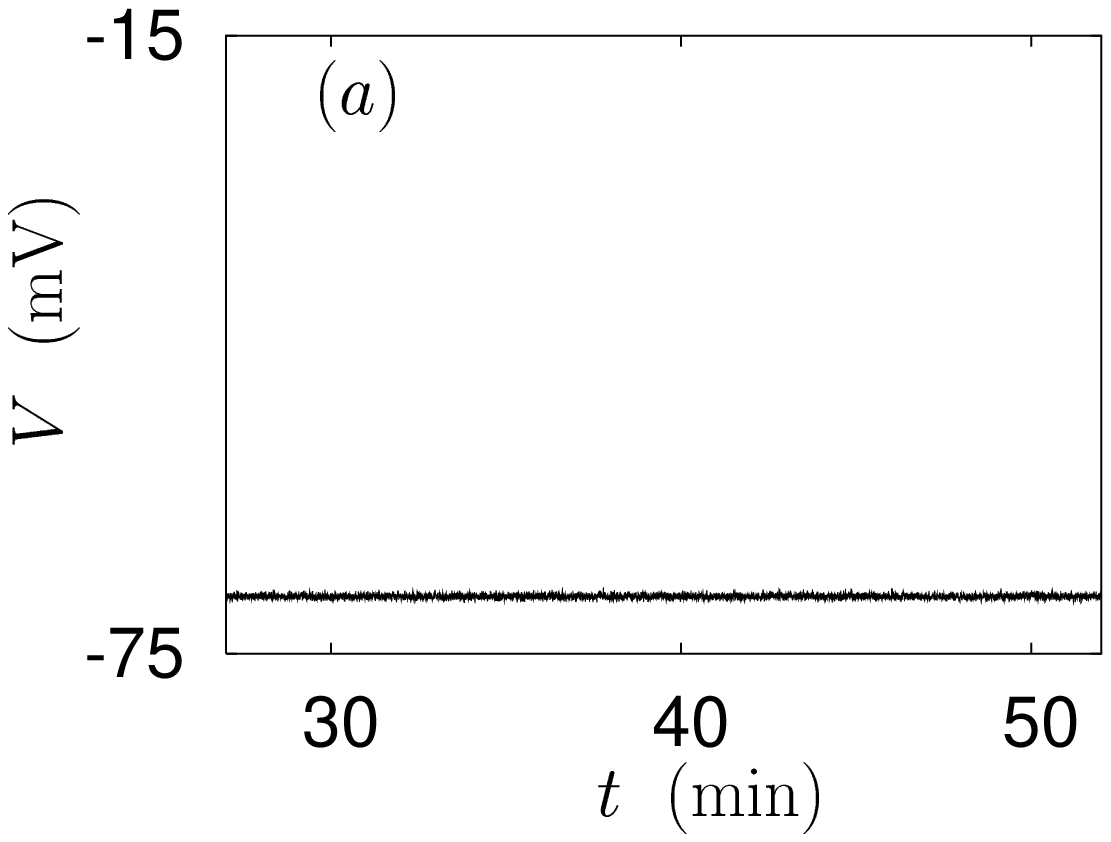} & 
\epsfig{width=4.2cm,file=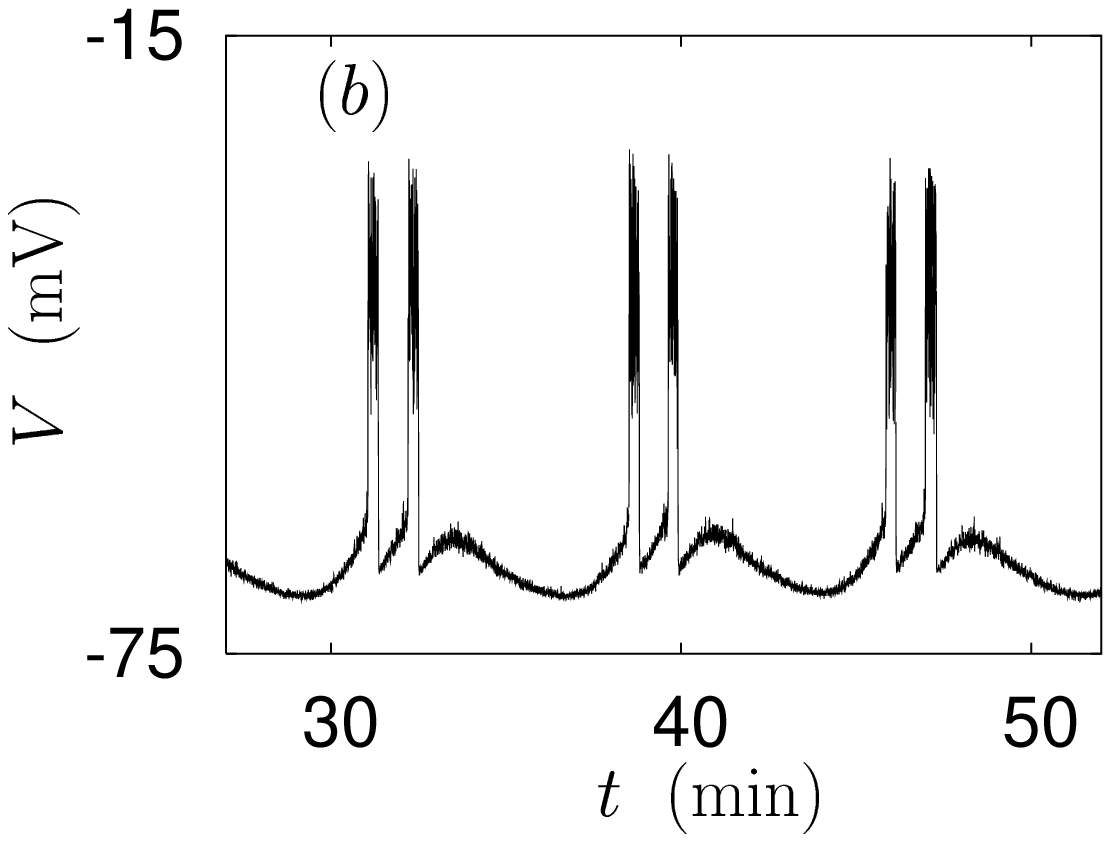} \\ 
\epsfig{width=4.2cm,file=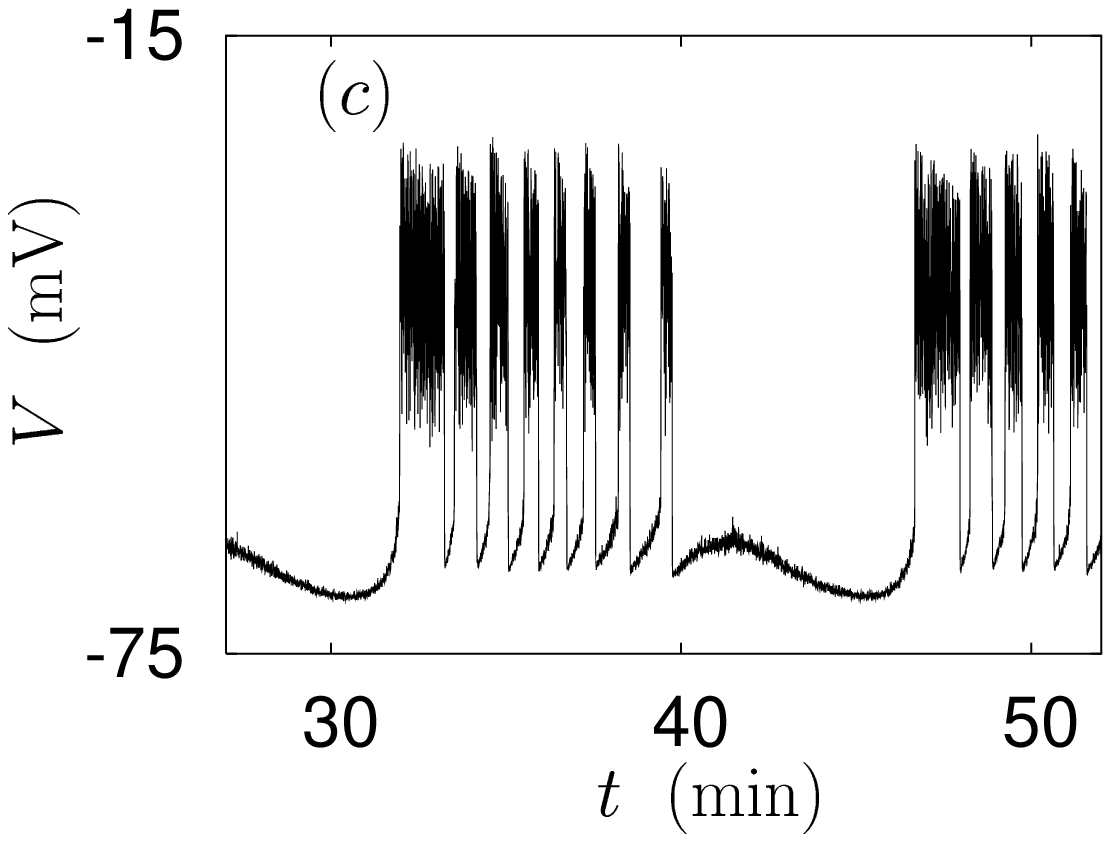} & 
\epsfig{width=4.2cm,file=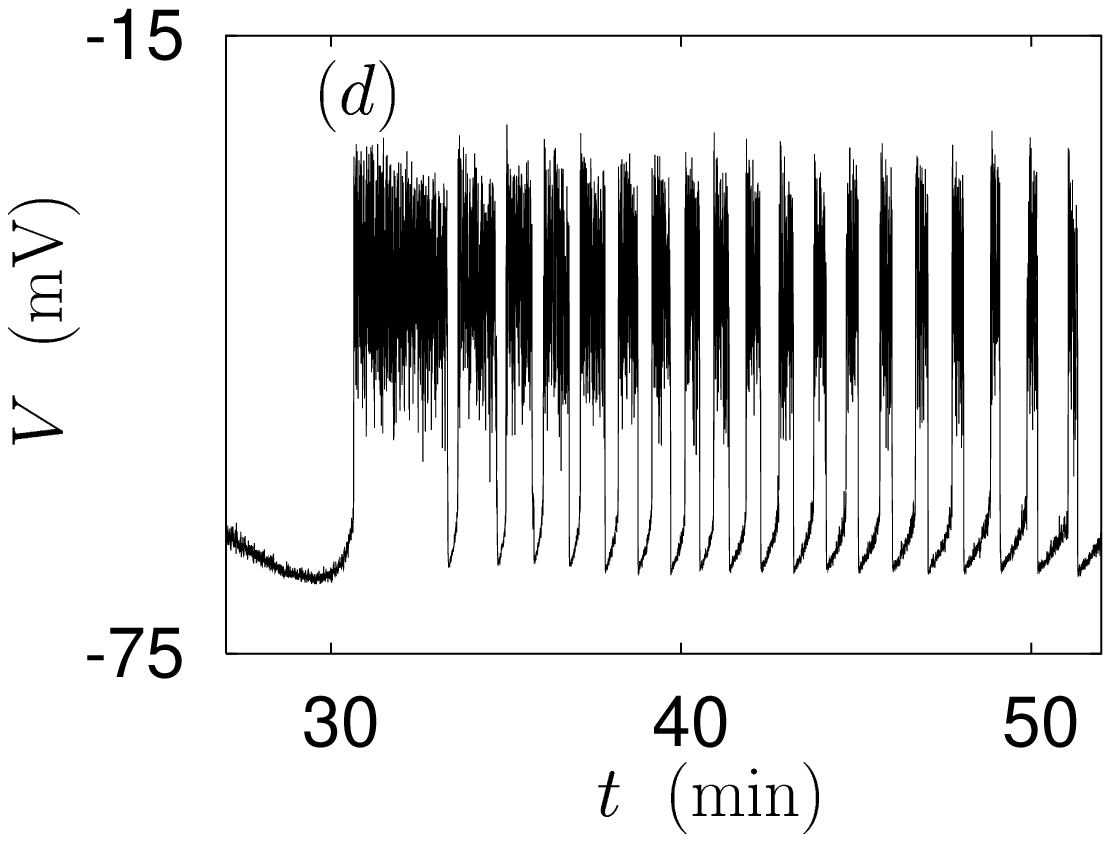} \\ 
\end{tabular} 
\caption[membrane potential]{ 
Bursting behavior of the membrane potential at several extracellular glucose concentrations: 
$G_0 = $\,(a)\,3, (b)\,5, (c)\,9, and (d)\,19\,mM.} 
\label{poten_s} 
\end{figure} 

\subsection{Indirect Pathway} 

\begin{figure} 
\epsfig{width=7.16cm,height=5.7cm,file=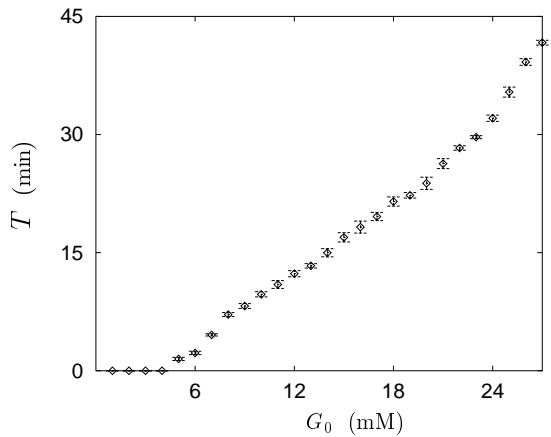} \\ 
\caption[Duration time $T$ of a bursting cluster depending on the extracellular 
glucose concentration] 
{Duration time $T$ of a bursting cluster depending on the extracellular glucose
concentration. Also shown is the typical error bar, estimated from the standard deviation.}
\label{duration} 
\end{figure}

\begin{figure} 
\begin{tabular}{cccc} 
\epsfig{width=4.2cm,file=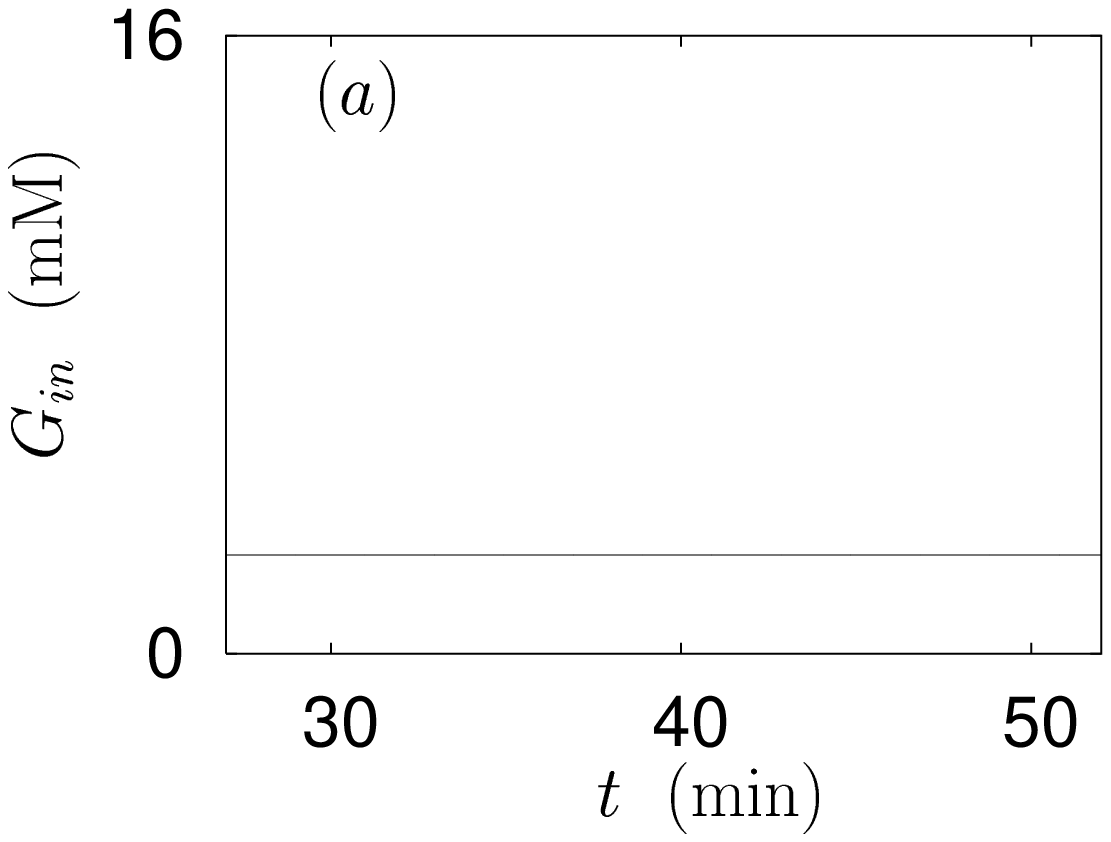} & 
\epsfig{width=4.2cm,file=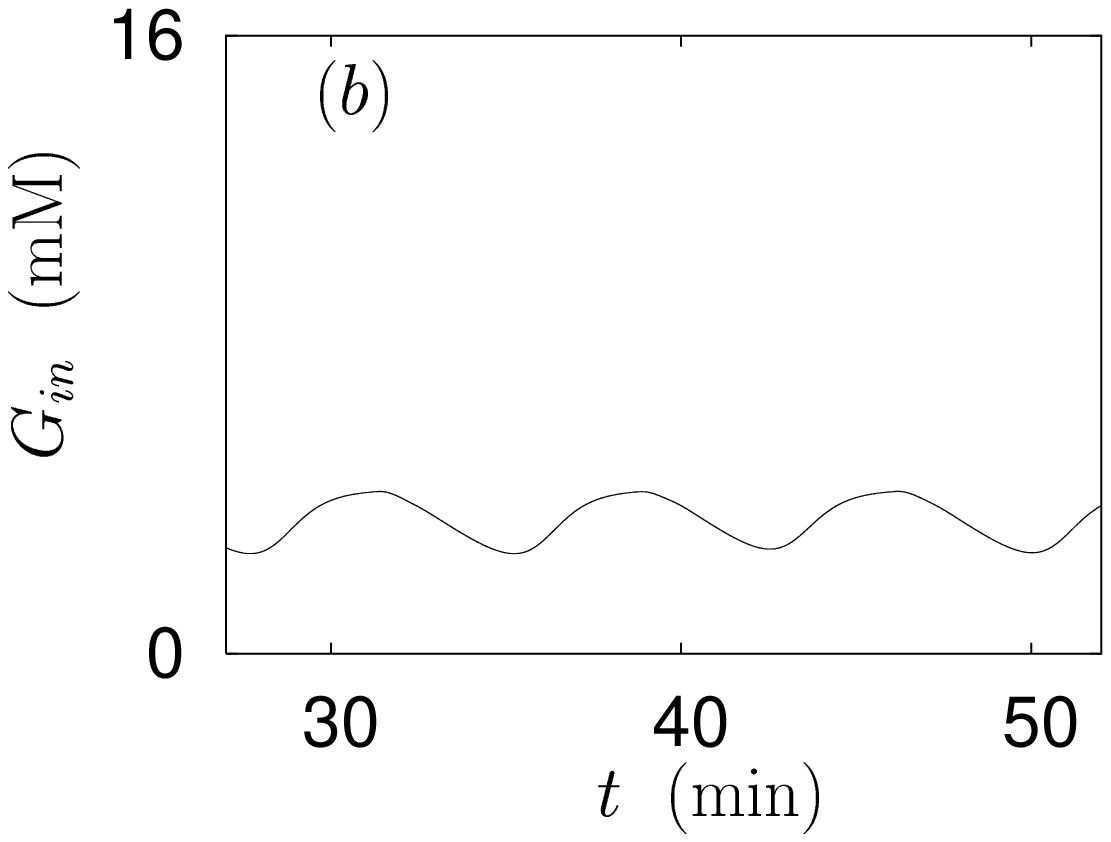} \\ 
\epsfig{width=4.2cm,file=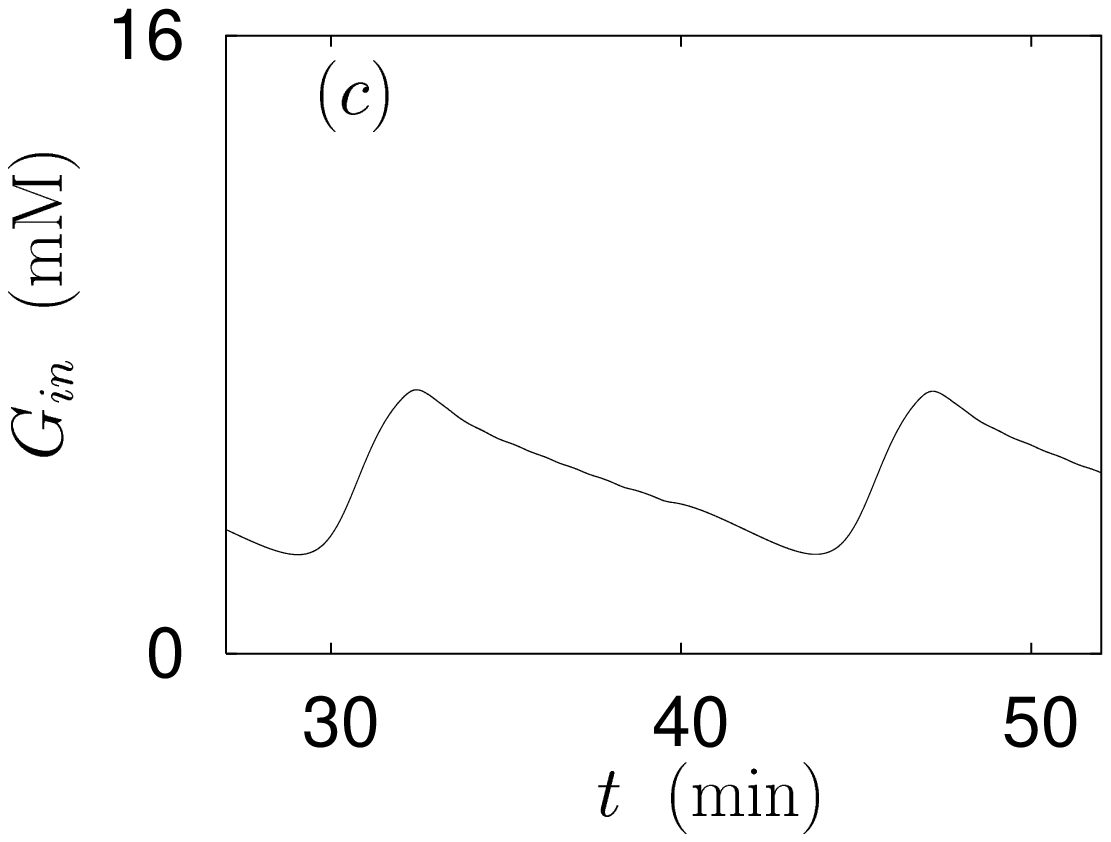} & 
\epsfig{width=4.2cm,file=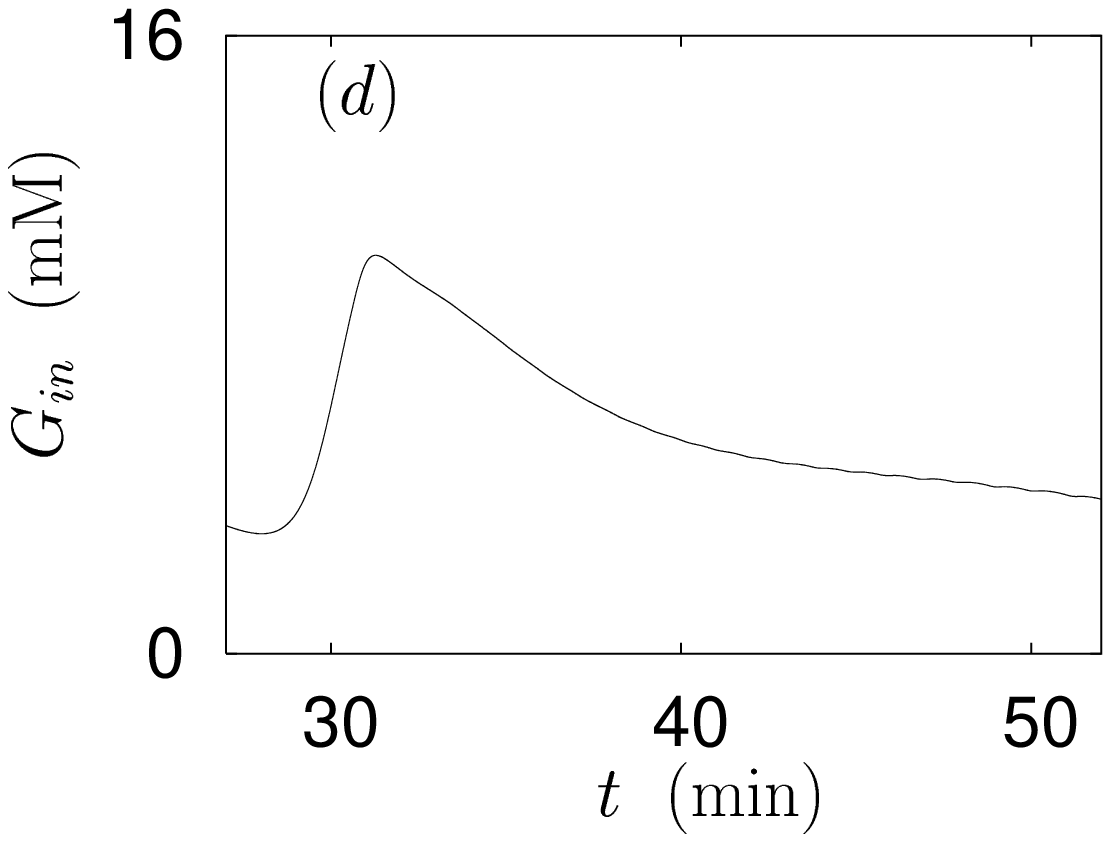} \\ 
\end{tabular} 
\caption[intracellular glucose concentration at several extracellular glucose concentrations]
{Behavior of the intracellular glucose concentration at the same extracellular 
glucose concentrations as in Fig. 3. 
} 
\label{Gi_s} 
\end{figure}

In order to consider indirect effects on $g_{K(ATP)}$ (via negative feedback to GLUT2 by
insulin), we take $f_D =1$ and $f_I = f$ with $J_0 = 0.35$, $\tau_J=1$\,min, and other
parameter values in Table I. 
In this manner we have performed simulations of an islet, which consists of $64$ cells
arranged on a two-dimensional square lattice shown in Fig.~\ref{lattice}, and obtained the
membrane potential, Ca$^{2+}$ concentration, and insulin secretion at the coupling
conductance $g_c = 0.06$. 
Figure~\ref{poten_s} shows the obtained time dependence of the membrane potential 
of a single cell randomly chosen among $64$ cells for extracellular glucose 
concentration $G_0 =$ (a) $3$, (b) $5$, (c) $9$, and (d) $19\,\rm{mM}$. 
Observed is the bursting behavior of the membrane potential, forming periodic clusters of
bursts. Also, the first burst persists longest, while following ones last 
progressively shorter, similar to experimental observations~\cite{Rorsman}. 
The total duration time of such a regularly bursting cluster grows with the glucose
concentration, as shown in Fig.~\ref{duration}. 
Here the criterion for an active phase has been chosen to be the membrane potential higher
than $V=-55\,\rm{mV}$; the duration time $T$ represents the total duration of such active
phases in a cluster, averaged over clusters. 
The intracellular glucose concentration $G_{in}$ of a single cell under the same conditions
as in Fig.~\ref{poten_s} is displayed in Fig.~\ref{Gi_s};
note the oscillatory behavior, with the amplitude and the period growing with the
extracellular glucose concentration $G_0$.  

\begin{figure}
\begin{tabular}{cccc}
\epsfig{width=4.2cm,height=3cm,file=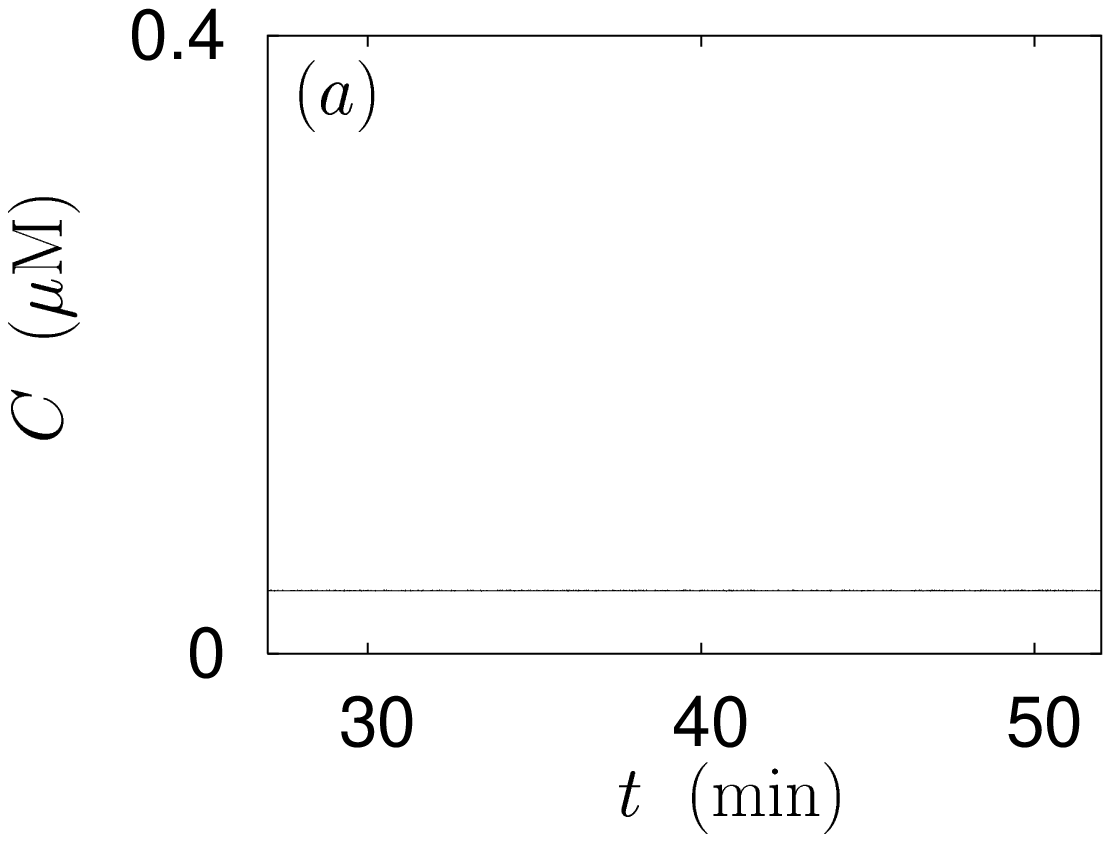} &
\epsfig{width=4.2cm,height=3cm,file=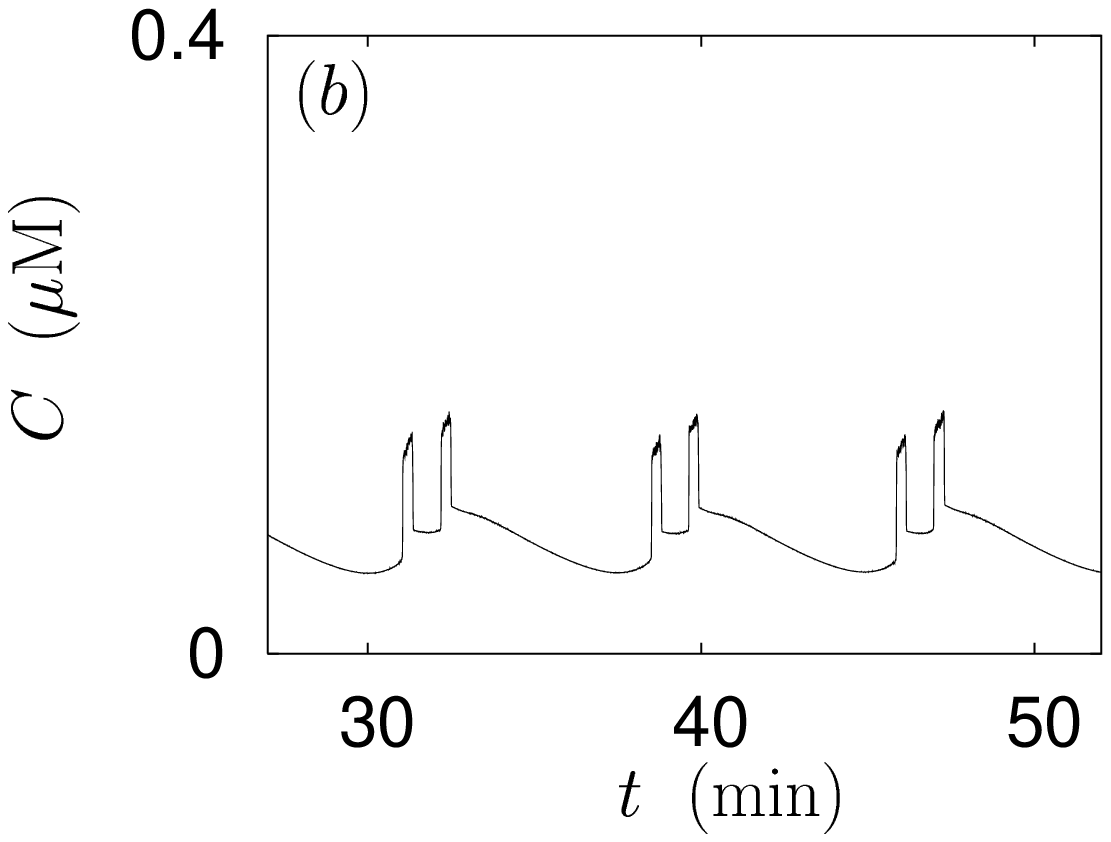} \\
\epsfig{width=4.2cm,height=3cm,file=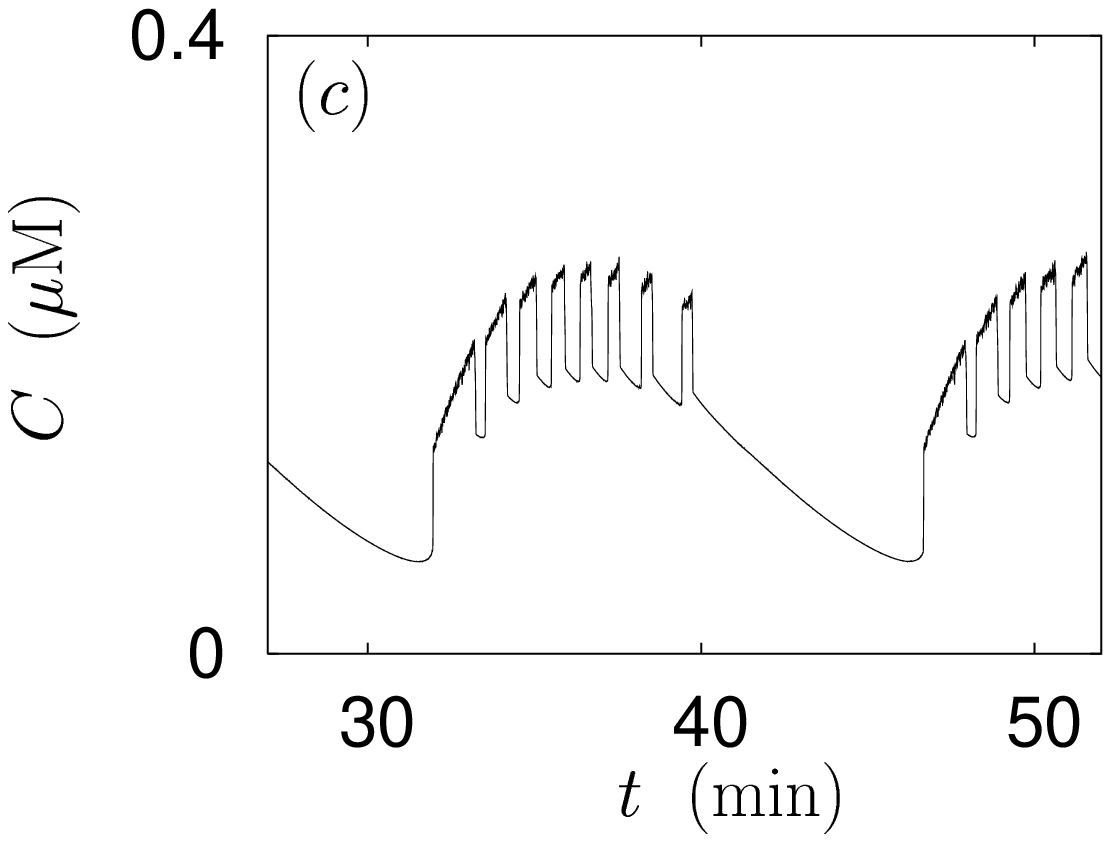} &
\epsfig{width=4.2cm,height=3cm,file=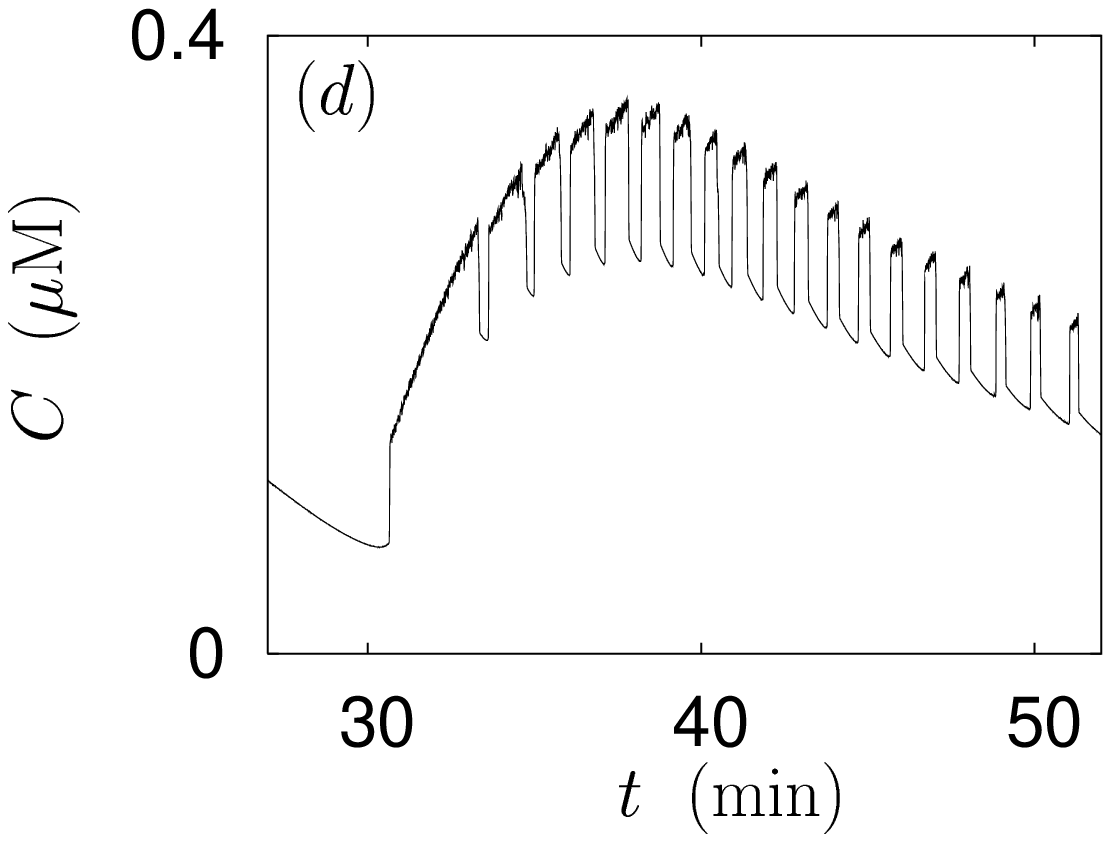} \\
\end{tabular}
\caption[Behavior of the calcium concentration at several extracellular glucose
concentrations]{Behavior of the calcium concentration at the same extracellular glucose
concentrations as in Fig. 3.
}
\label{cal_s}
\end{figure}

\begin{figure}
\begin{tabular}{cccc}
\epsfig{width=4.2cm,height=3cm,file=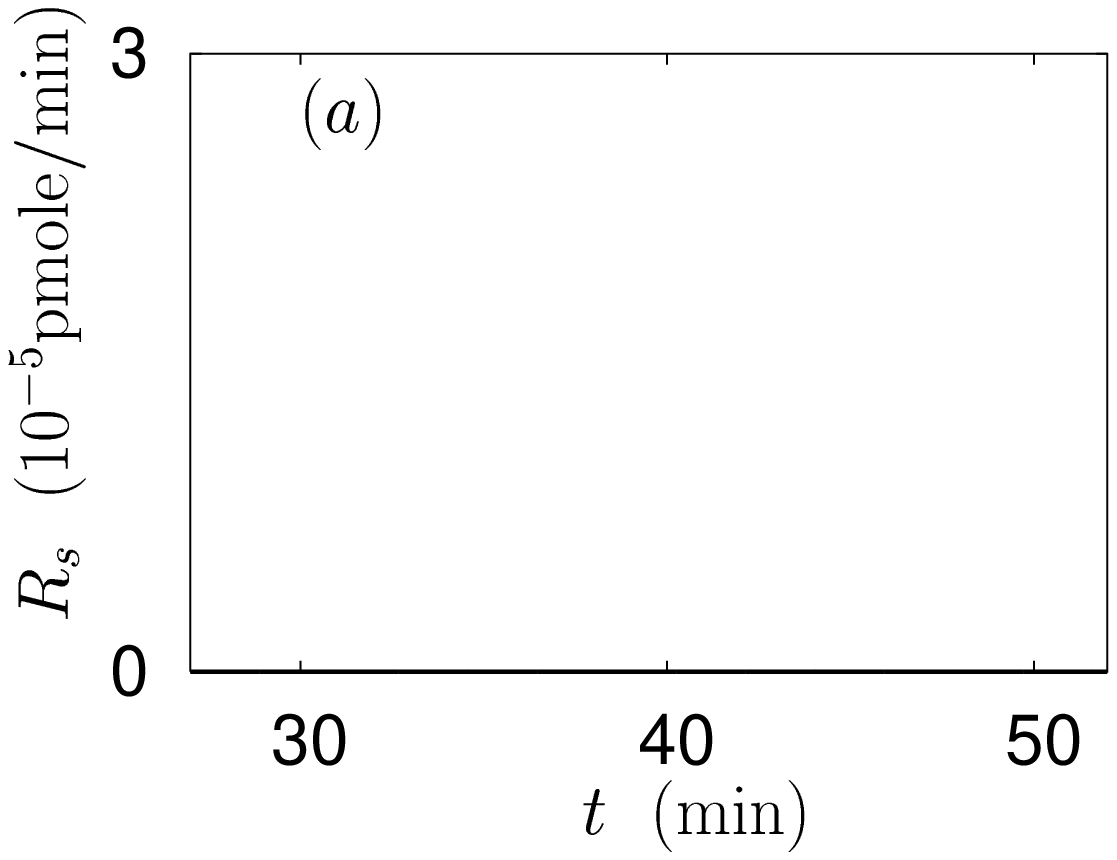} &
\epsfig{width=4.2cm,height=3cm,file=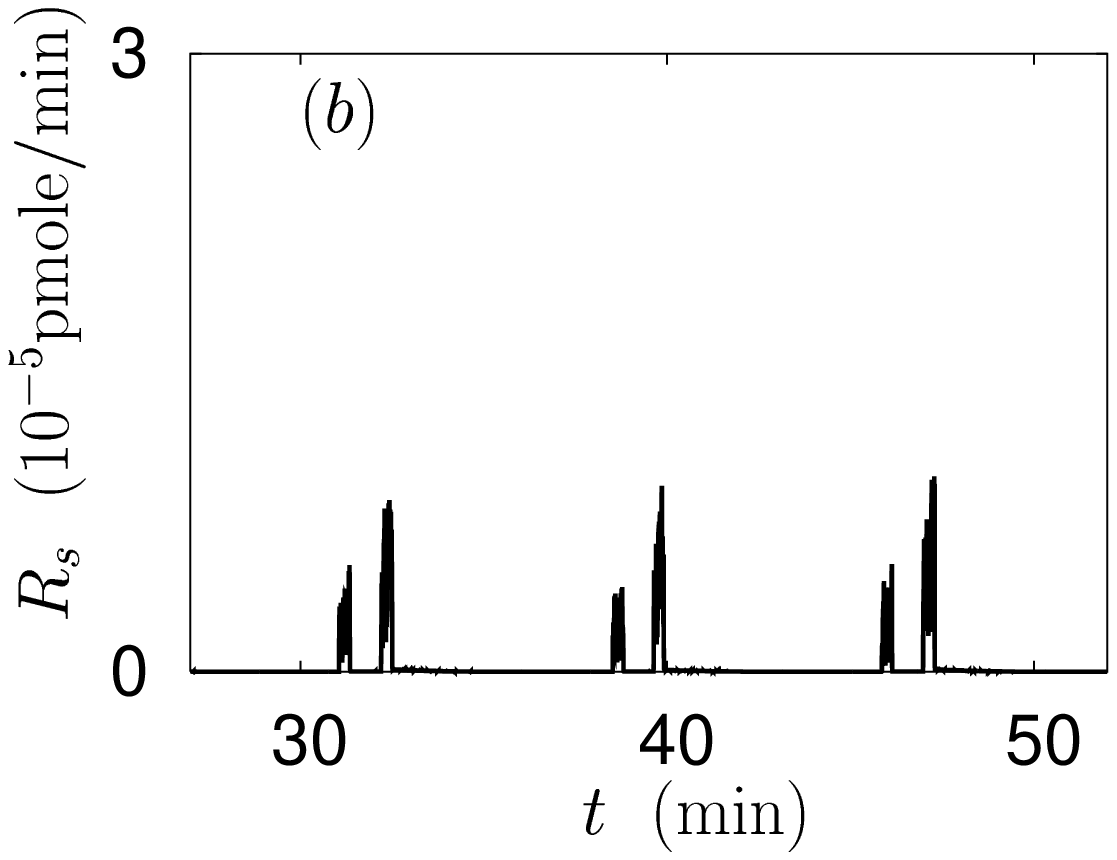} \\
\epsfig{width=4.2cm,height=3cm,file=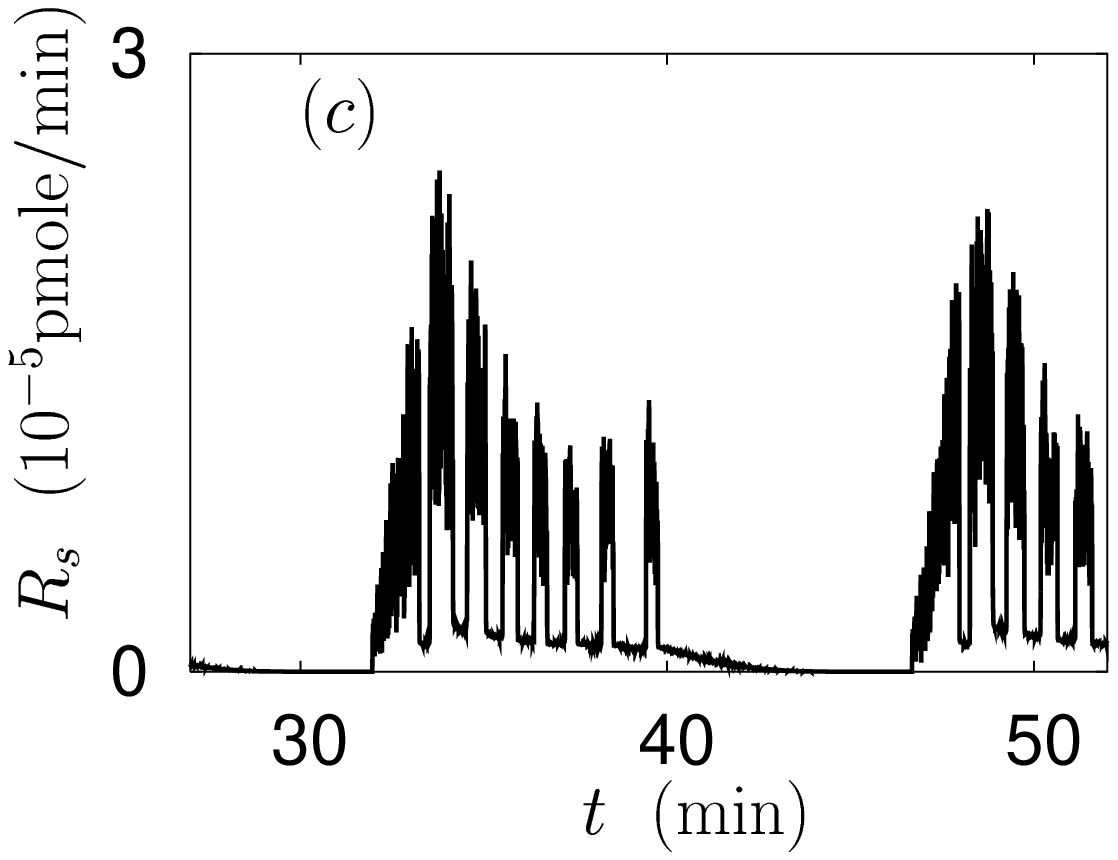} & 
\epsfig{width=4.2cm,height=3cm,file=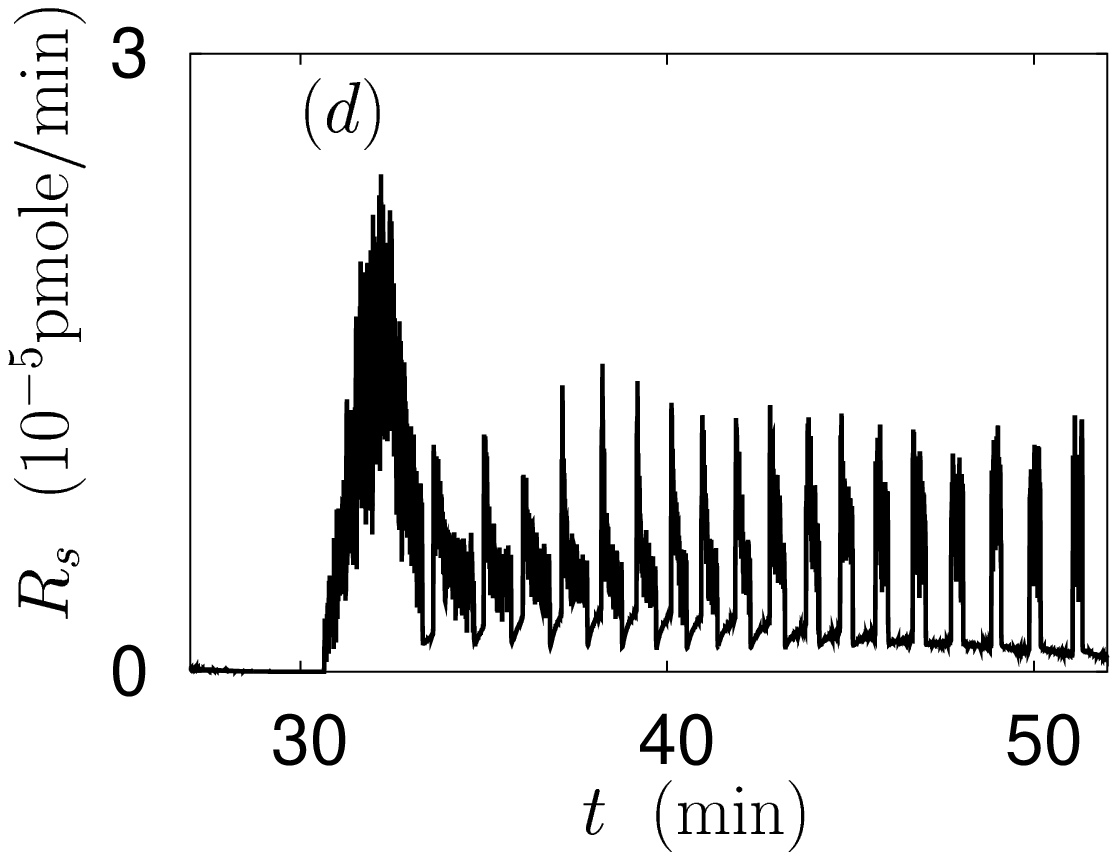} \\ 
\end{tabular} 
\caption[Time dependence of the insulin secretion rate at several extracellular glucose
concentrations]{Behavior of the insulin secretion rate at the same extracellular glucose
concentrations as in Fig. 3. 
} 
\label{ISR_s} 
\end{figure}

\begin{figure} 
\begin{tabular}{cccc}
\epsfig{width=4.2cm,height=3cm,file=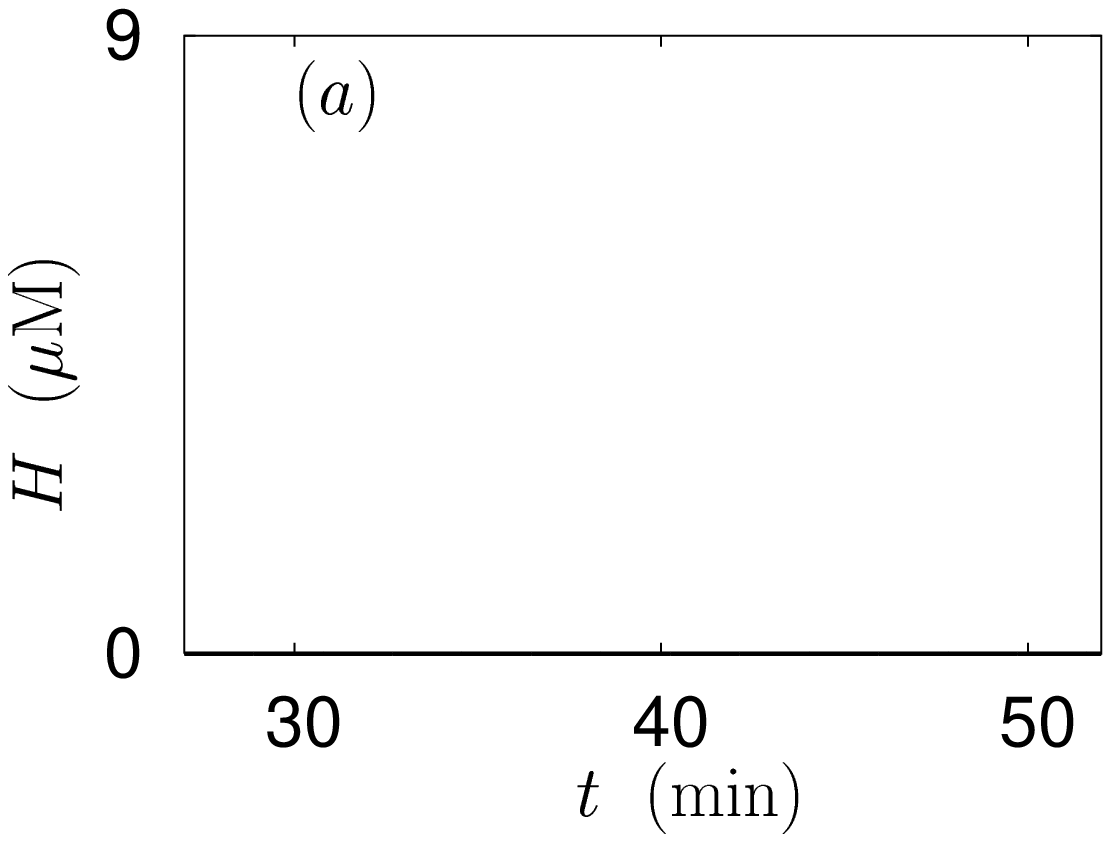} & 
\epsfig{width=4.2cm,height=3cm,file=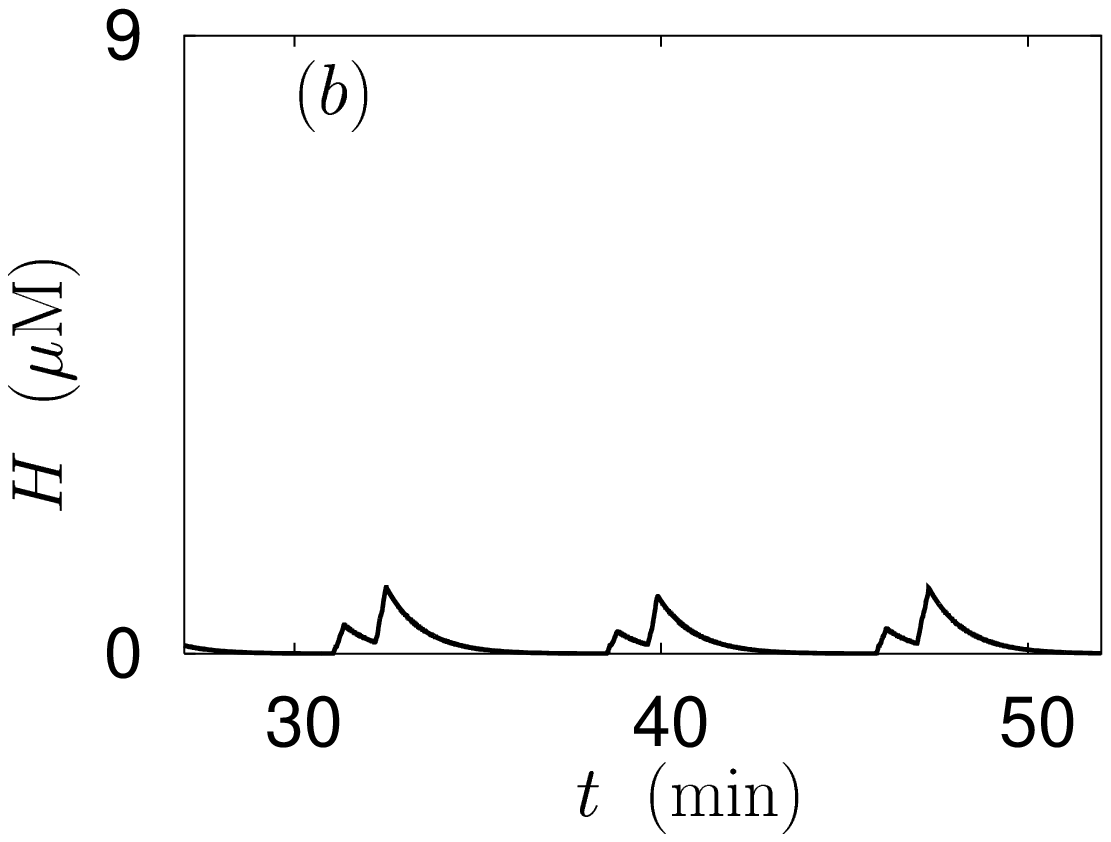} \\ 
\epsfig{width=4.2cm,height=3cm,file=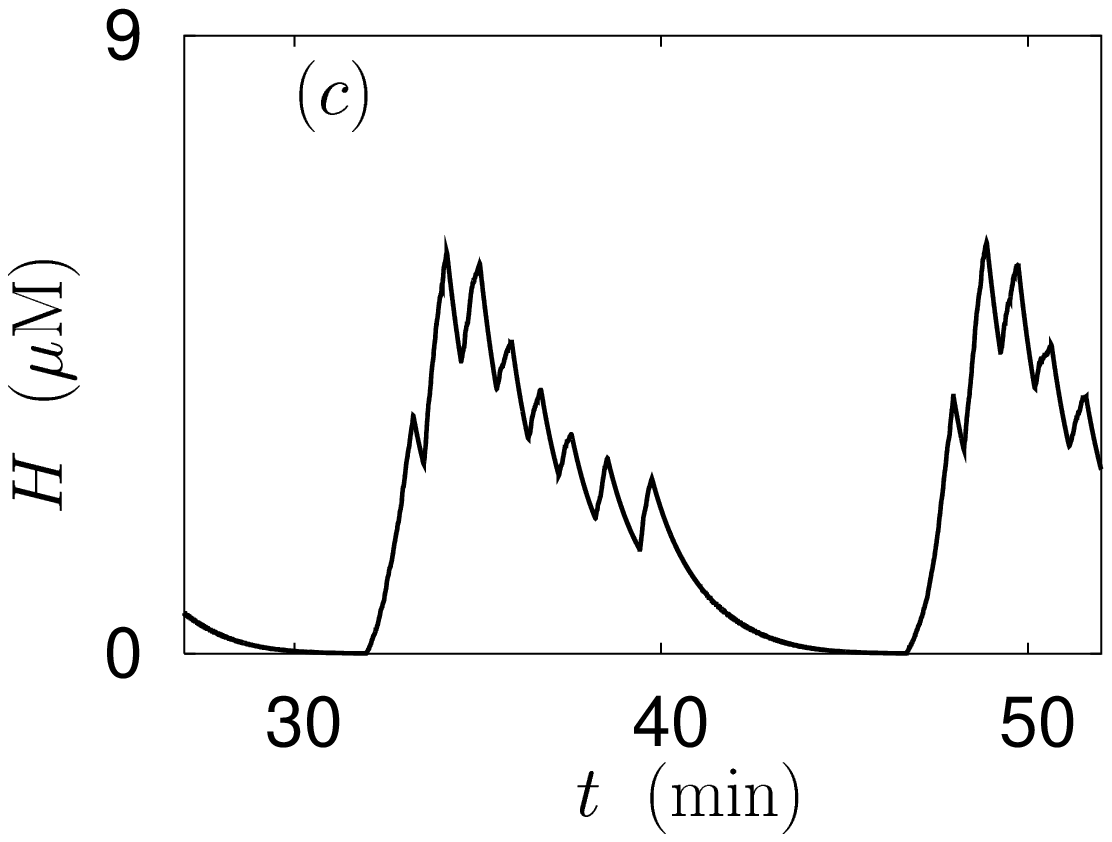} & 
\epsfig{width=4.2cm,height=3cm,file=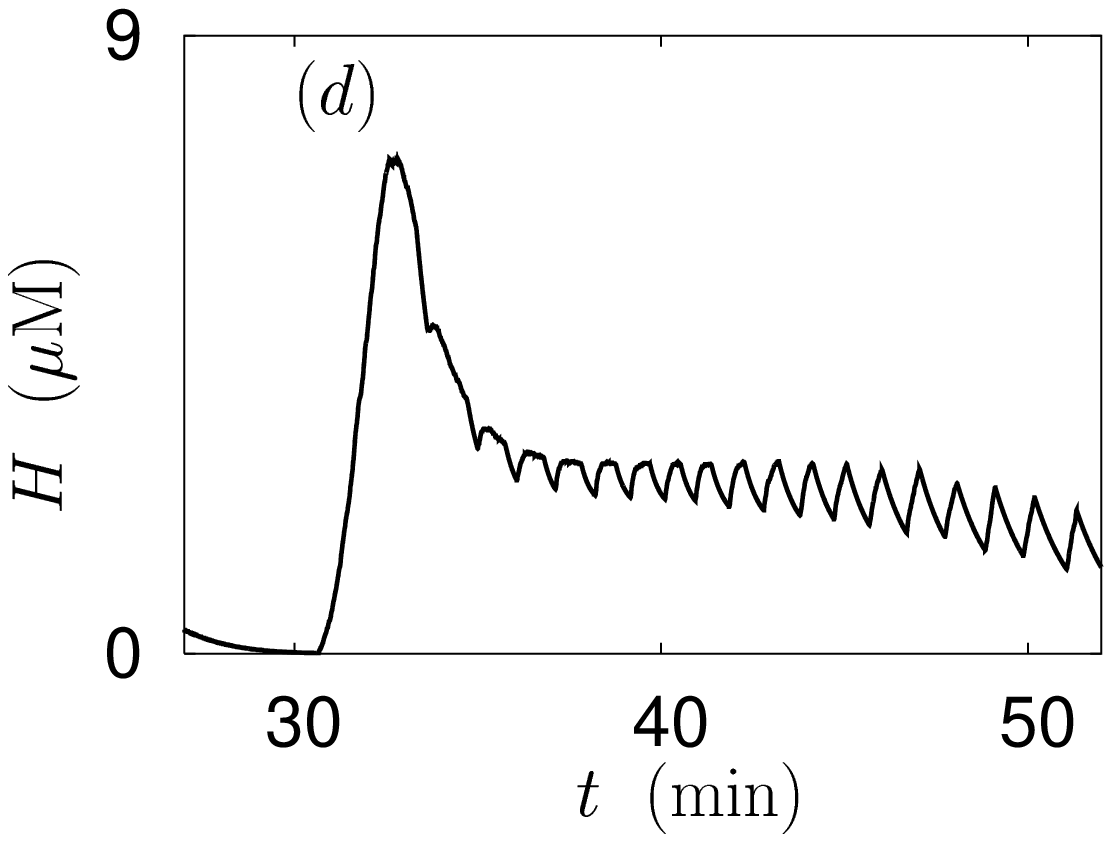} \\ 
\end{tabular} 
\caption[Behavior of the insulin concentration at several extracellular glucose
concentrations]{Behavior of the insulin concentration around a cell at the same
extracellular glucose concentrations as in Fig. 3. 
} 
\label{Ins_s} 
\end{figure}

Figures~\ref{cal_s}, ~\ref{ISR_s}, and ~\ref{Ins_s} exhibit the corresponding behaviors of
the concentration of cytosolic calcium Ca$^{2+}$, the insulin secretion rate, and the insulin
concentration, respectively, under the same glucose concentrations as in Fig.~\ref{poten_s}. 
Unlike other quantities obtained for a single cell, the calcium concentration in
Fig.~\ref{cal_s} has been averaged over all cells. 
Note that oscillations of the averaged calcium concentration have periods similar to those of
repetitive activation of the membrane potential;
this reflects that most of the cells in an islet are well synchronized except for
slight phase shift~\cite{synchrony}, as observed in simulations.
It is observed that the Ca$^{2+}$ concentration keeps increasing during the action potential
firing and then decreases slowly after the firing stops.
The insulin release occurs in the form of sharp peaks as soon as the Ca$^{2+}$ concentration
increases in the cell (see Fig.~\ref{ISR_s}).
After the rest period of about 5\,min, the secretion rate of insulin reaches the maximum
value and then diminishes at the next burst, as the amount of insulin stored for
rapid secretion reduces. The overall behavior of the insulin secretion rate agrees
well with the experimental results~\cite{Gilon}.

\begin{figure} 
\begin{tabular}{cccc} 
\epsfig{width=4.2cm,height=3cm,file=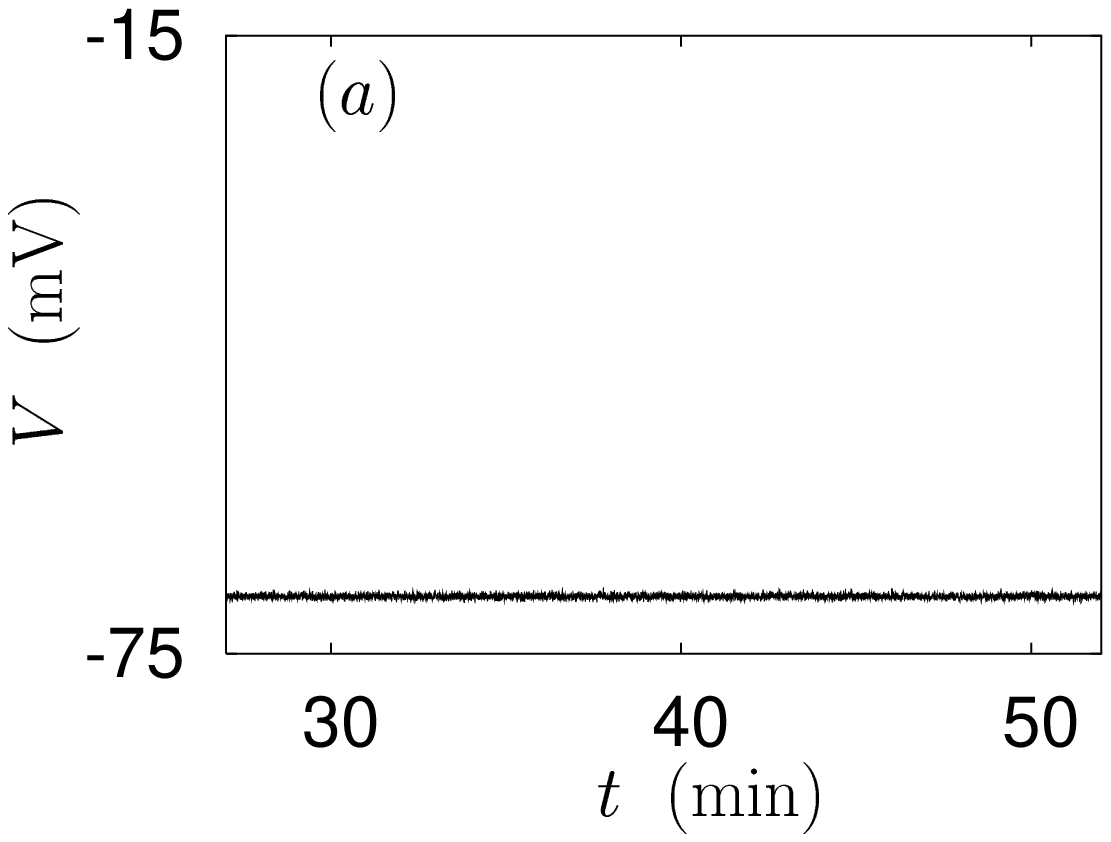} & 
\epsfig{width=4.2cm,height=3cm,file=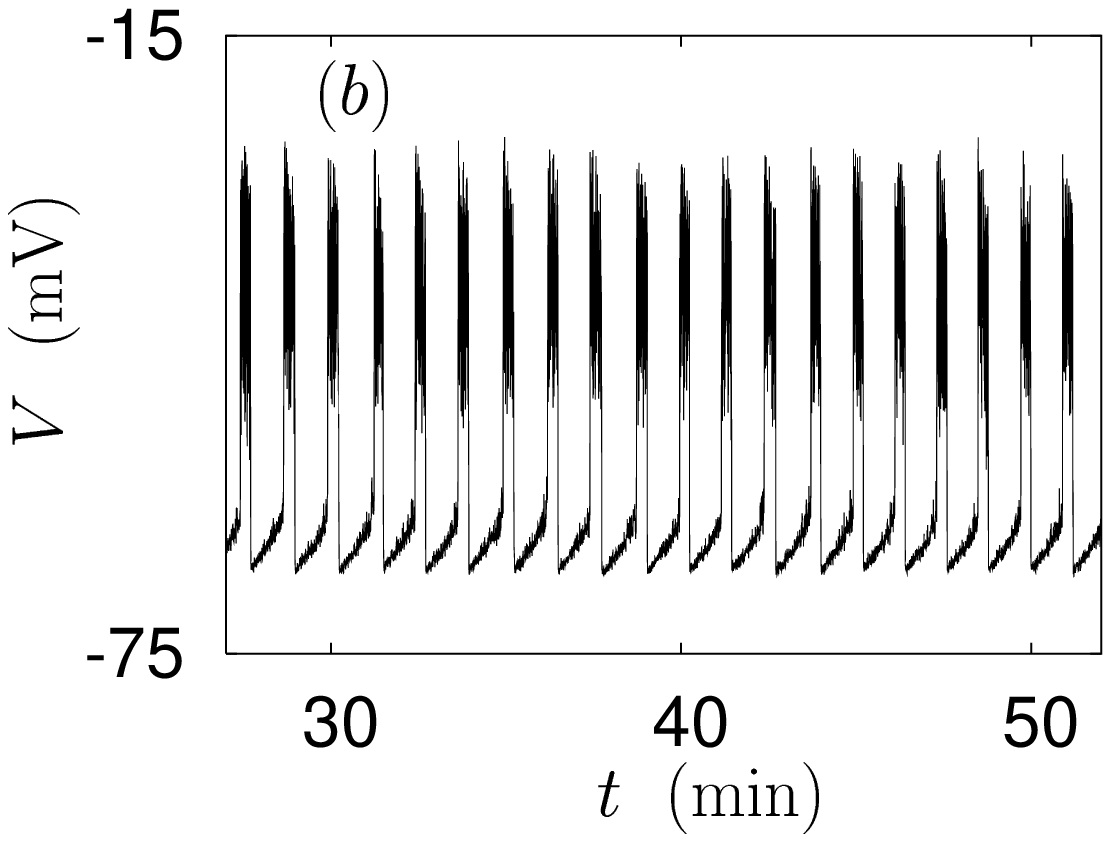} \\ 
\epsfig{width=4.2cm,height=3cm,file=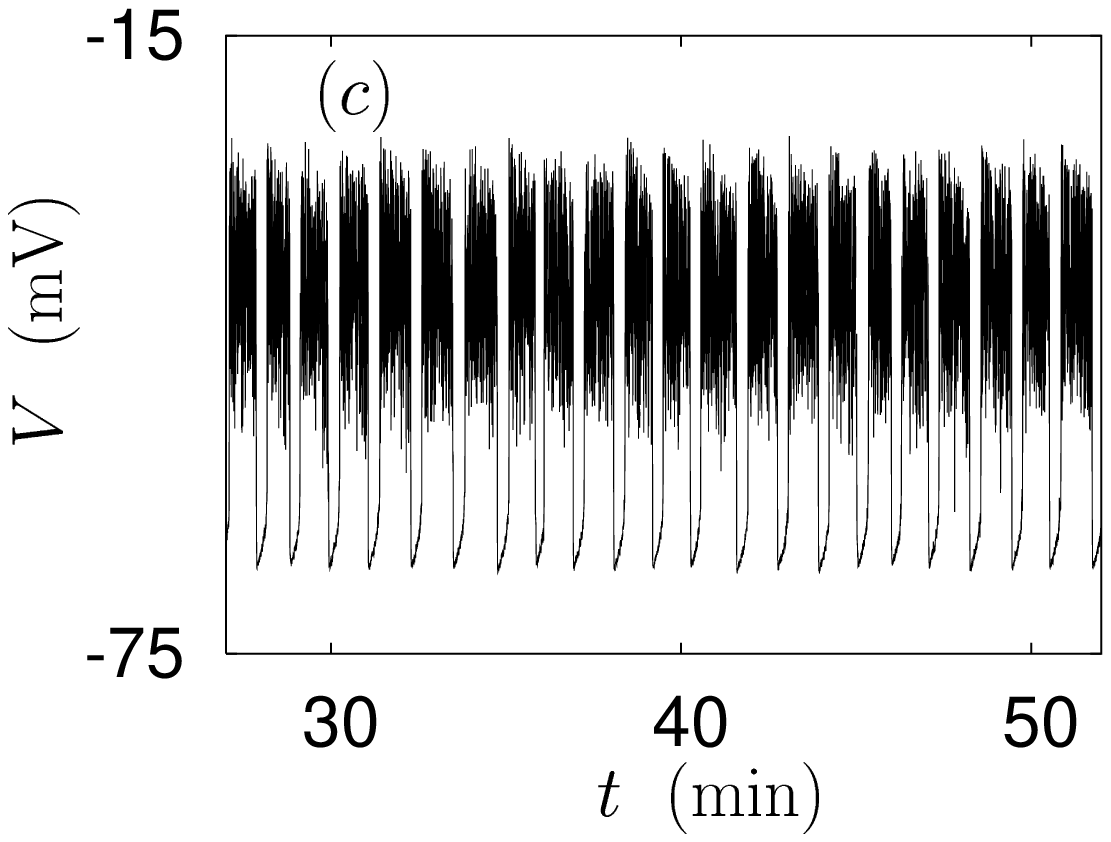} & 
\epsfig{width=4.2cm,height=3cm,file=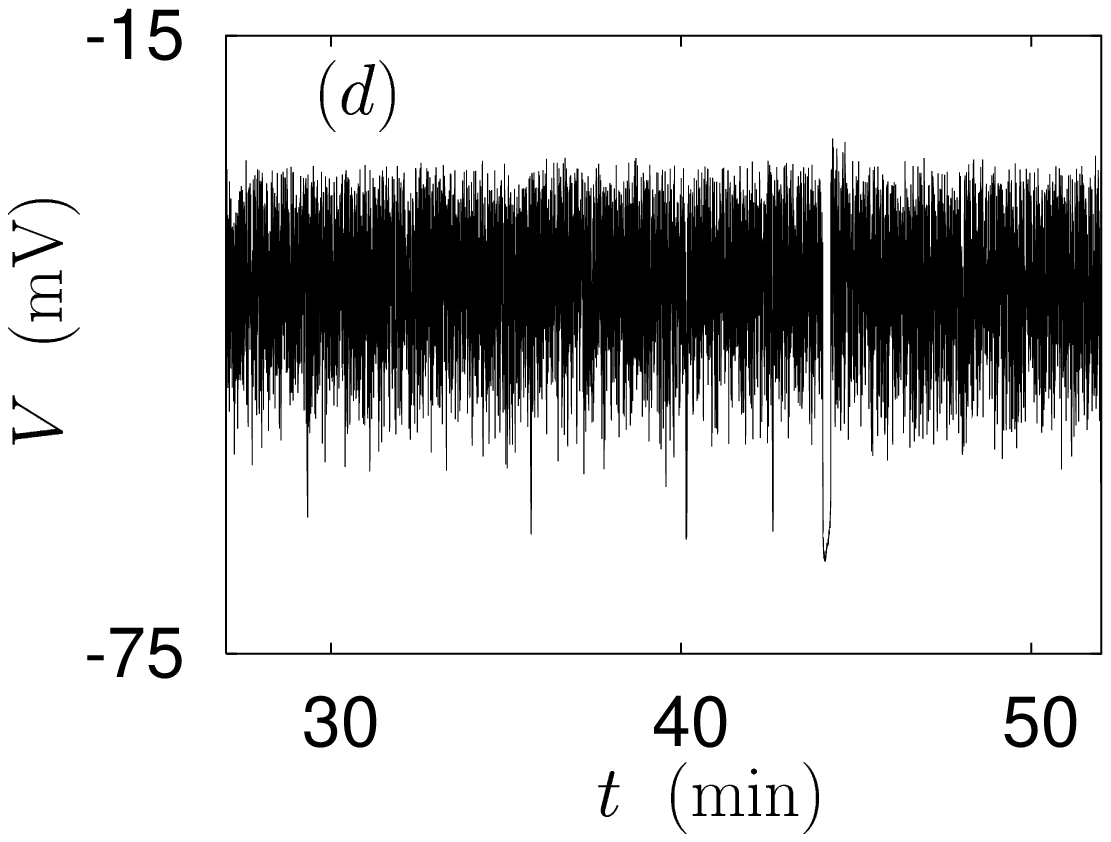} \\ 
\end{tabular} 
\caption[Membrane potential in the stationary]{ 
Behavior of the membrane potential in the stationary state of the intracellular glucose
concentration. The extracellular glucose concentrations are the same as those in Fig. 3. 
} 
\label{steady} 
\end{figure}

When the flow rate is sufficiently high, on the other hand, the intracellular glucose 
concentration tends to saturate rather than to oscillate. 
To probe bursting patterns in this stationary state, we set the insulin flow rate
$k_0=9.2\,\rm{s}^{-1}$, maintaining previous values of other parameters. 
Such a low insulin feedback condition may correspond to the experiment performed to obtain
stationary values of the intracellular glucose concentration at several extracellular
glucose concentrations~\cite{Whitesell}.
The resulting behavior of the membrane potential is displayed in Fig.~\ref{steady}, 
depending on the extracellular glucose concentration.
The duration time of each burst is also observed to grow with the extracellular glucose 
concentration; on the other hand, oscillatory patterns with periods of about 10\,min do not
emerge. 
It is thus concluded that the emergence of slow oscillations of the membrane potential 
depends upon the flow rate of the circulating extracellular fluid, i.e., blood. 
The lack of slow oscillations led by minimal insulin feedback matches well with the
experimental results of IRS-1 (insulin receptor substrate 1) knockout mice~\cite{Kulkarni}. 

\begin{figure} 
\epsfig{width=7.16cm,height=5.7cm,file=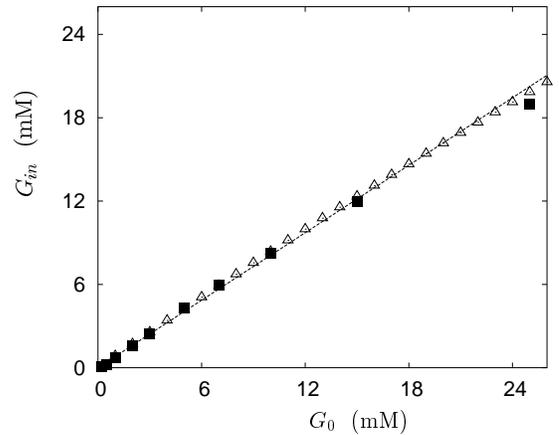} \\
\caption[Saturated intracellular glucose concentration vs the extracellular glucose
concentration]{Saturated intracellular glucose concentration depending on the extracellular
glucose concentration. Squares represent the data obtained from experiment~\cite{Whitesell},
whereas triangles are the data obtained via simulations and fitted by the dashed line. 
} 
\label{G0vsGi} 
\end{figure}

Figure~\ref{G0vsGi} shows how the intracellular glucose concentration $G_{in}$ depends on
the extracellular glucose concentration $G_0$. Shown together with the experimental 
results~\cite{Whitesell} are the saturated values of $G_{in}$ obtained from simulations 
with $G_0$ varied from $1$\,mM to $26$\,mM. It is pleasing to see the excellent 
agreement between the experimental data and the simulation data. 
In particular, the dependence of $G_{in}$ on $G_0$ yields reasonably good linear fitting, 
with the slope $0.81$. 

\begin{figure}
\begin{tabular}{cccc}
\epsfig{width=7.16cm,file=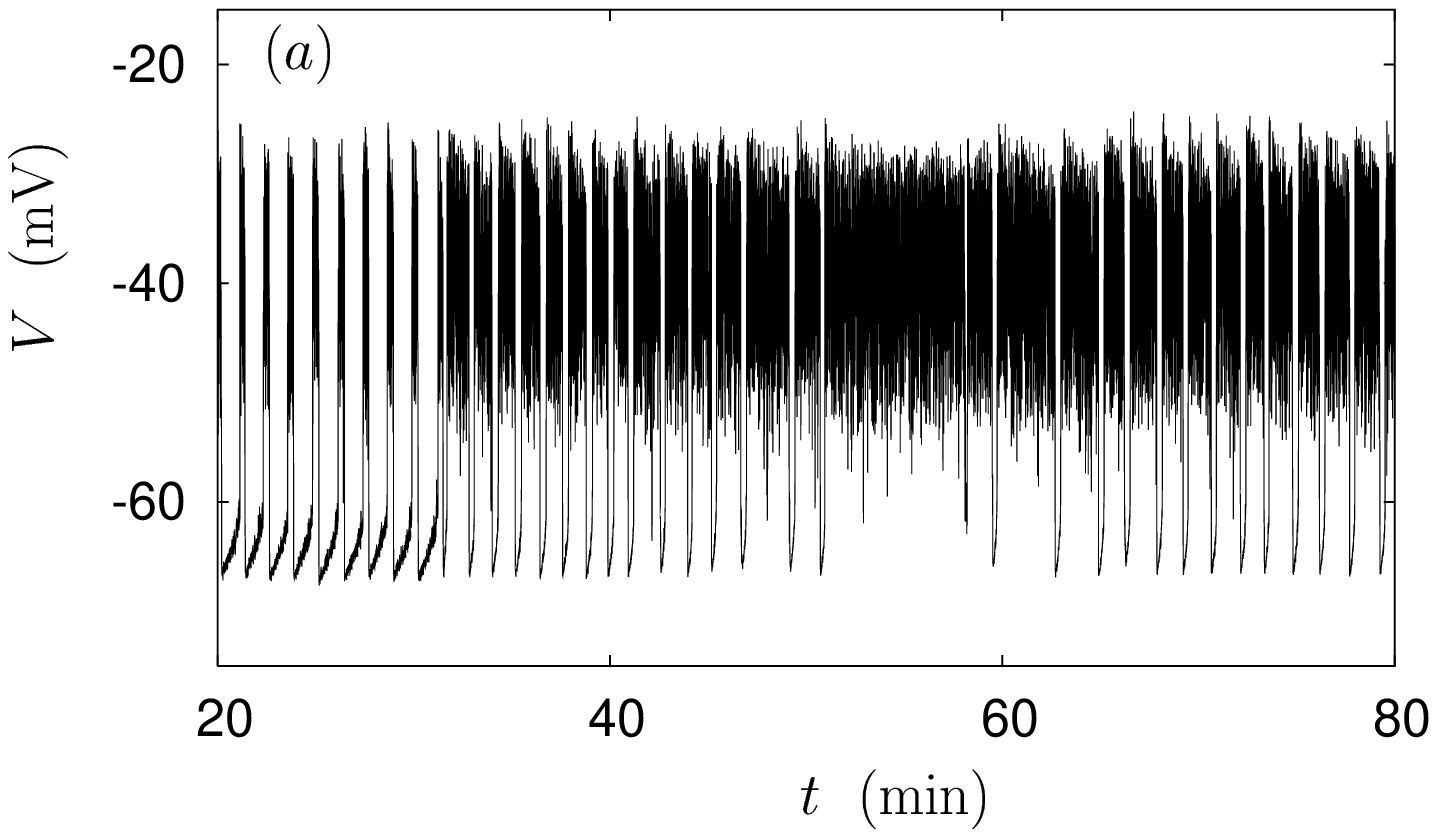} \\
\epsfig{width=7.16cm,file=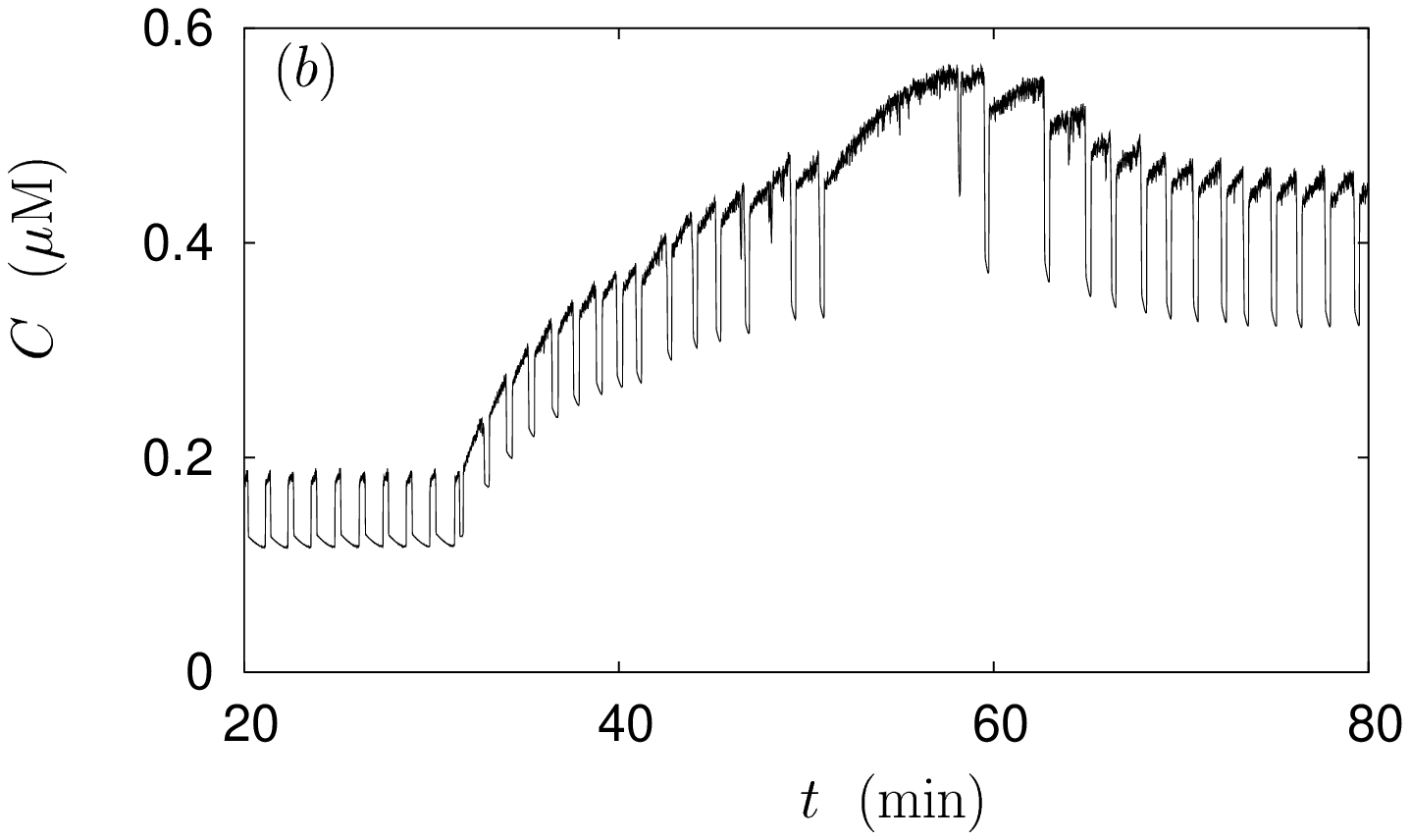} \\
\epsfig{width=7.16cm,file=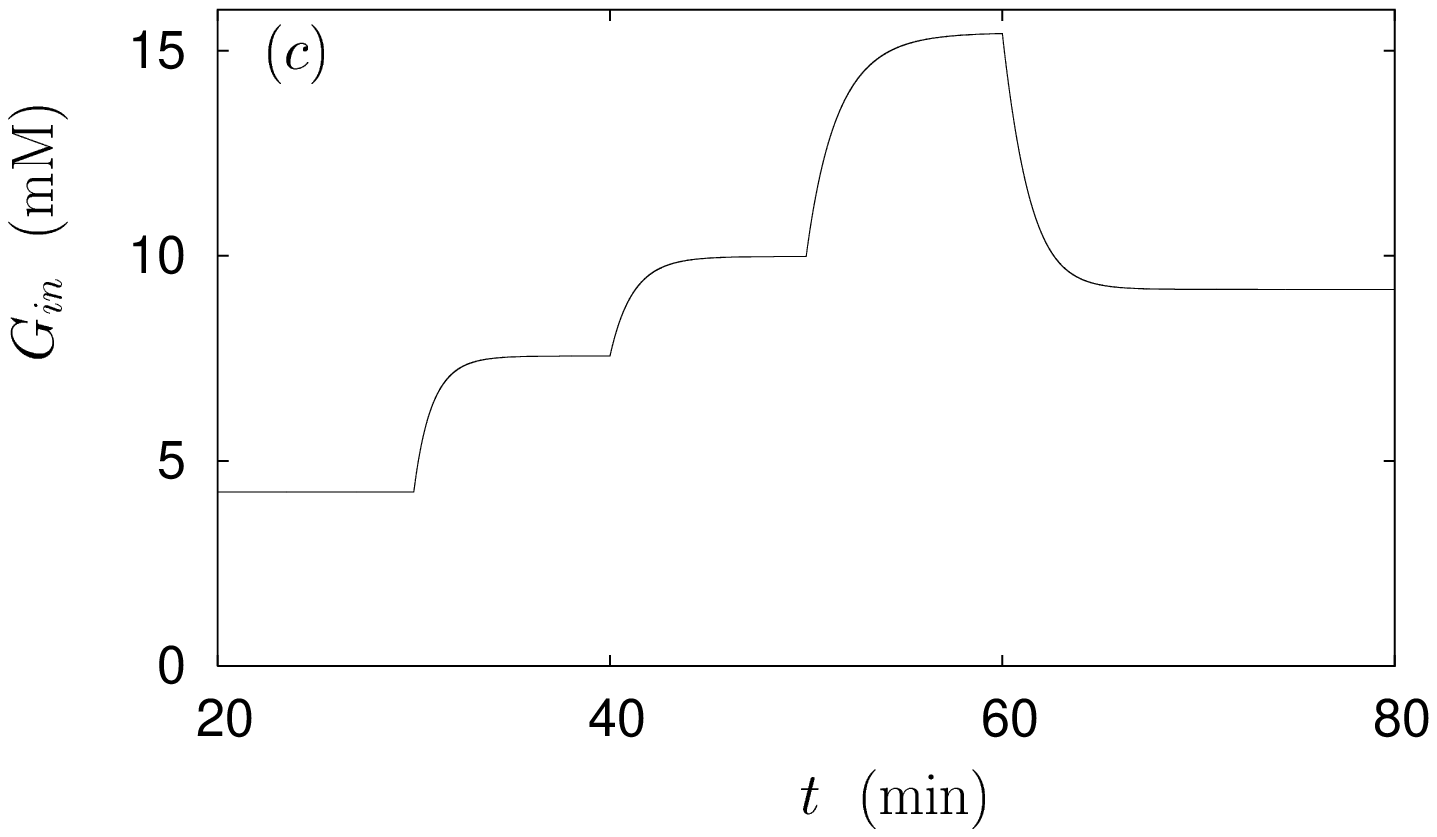} \\
\end{tabular}
\caption{
Behavior of (a) the membrane potential, (b) calcium concentration,
and (c) intracellular glucose concentration when glucose is injected as a step function
of time: $G_0 = 5, \,9, \,12, \,19$, and $11\,$mM at time $t = 20, \,30, \,40, \,50$,
and $60$\,min, respectively.
}
\label{inject}
\end{figure}

We also investigate the case that the concentration of injected glucose is varied 
(with $k_0 = 9.2\,$s$^{-1}$), to probe time-dependent responses to sudden changes of the
extracellular glucose concentration. 
For simplicity, the extracellular glucose concentration $G_0$ is given as a step function of
time, with possible time delay disregarded. 
Figure~\ref{inject} shows the behaviors of the membrane potential, calcium concentration,
and intracellular glucose concentration as glucose is injected with the concentration 
$G_0 =5, 9, 12, 19$, and $11$\,mM at time $t =20, 30, 40, 50$, and $60$\,min, respectively. 
They are observed to display behaviors in close response to the change of the glucose
concentration.

\begin{figure}
\centering
\epsfig{width=7.16cm,file=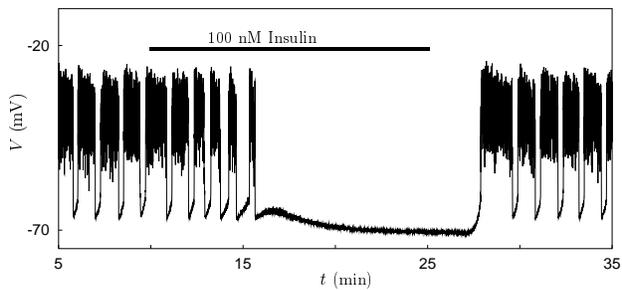} \\
\caption{Behavior of the membrane potential under $G_0=10$\,mM, when insulin corresponding
to $H_0=100$\,nM is injected during the time $t=10$\,min to $25$\,min.
The experimental values $c_1=2.9\times10^{-4}$M/s and $k_0 = 9.2\,$s$^{-1}$ are used.
}
\label{burstih}
\end{figure}

To confirm the assertion that insulin inhibits bursting of membrane
potentials, we examine the behavior of the membrane potential in response to the
injection of insulin.
Assuming that the basal plasma insulin concentration changes rapidly, we represent
the insulin stimulus by an increase in the basal insulin concentration $H_0$.
Figure~\ref{burstih} shows the membrane potential in response to the
insulin injection corresponding to $H_0=100$\,nM during the time $t = 10$\,min to $25$\,min
under the glucose concentration $G_0 =10$\,mM.
(Note that the parameter values for normal $\beta$ cells are used here for comparison,
different from those for hamster insulinoma tumor cells in Table I.
See Ref.~\onlinecite{Maki} and references therein.)
It is observed that the bursting action potential disappears after about $6$\,min
from the beginning of the insulin injection and appears again in about $3$\,min
after the cease of injection.
This is consistent with the experimental result~\cite{Khan}.

Finally, for a more realistic model, we allow variations in some parameters among the cells,
and examine the effects of such inhomogeneity. We have considered variations in the coupling
conductance $g_c$, choosing randomly its value in the range $0.05< g_c <0.07$.
The resulting behavior of the membrane potential is found to be qualitatively the same
as that for constant $g_c$.

\subsection{Direct Pathway}

\begin{figure}
\begin{tabular}{cccc}
\epsfig{width=4.2cm,height=3cm,file=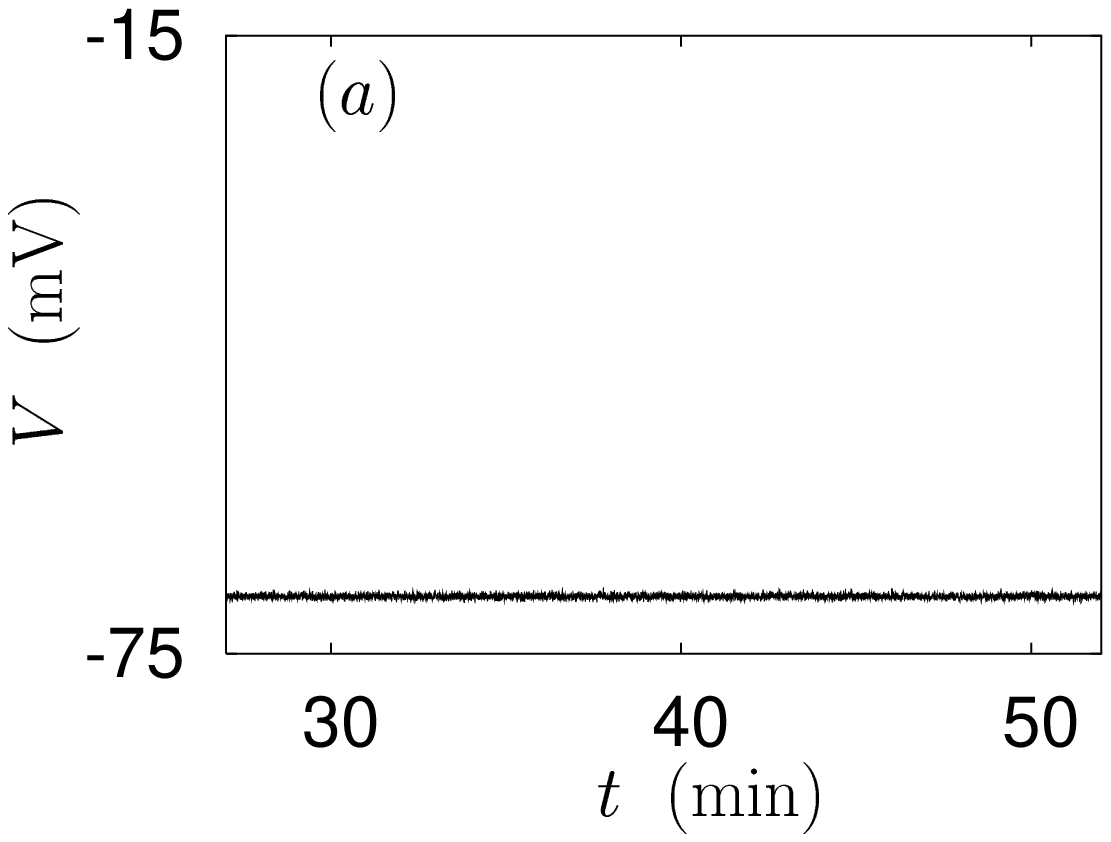} &
\epsfig{width=4.2cm,height=3cm,file=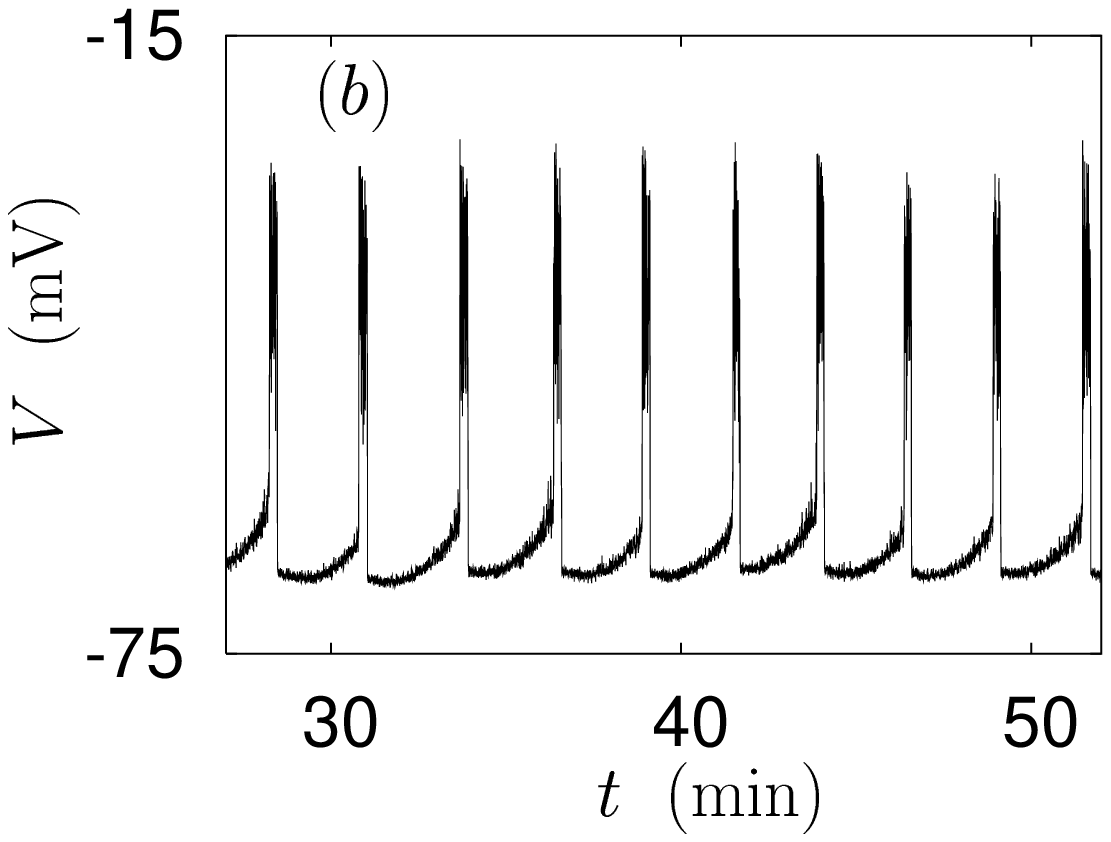} \\
\epsfig{width=4.2cm,height=3cm,file=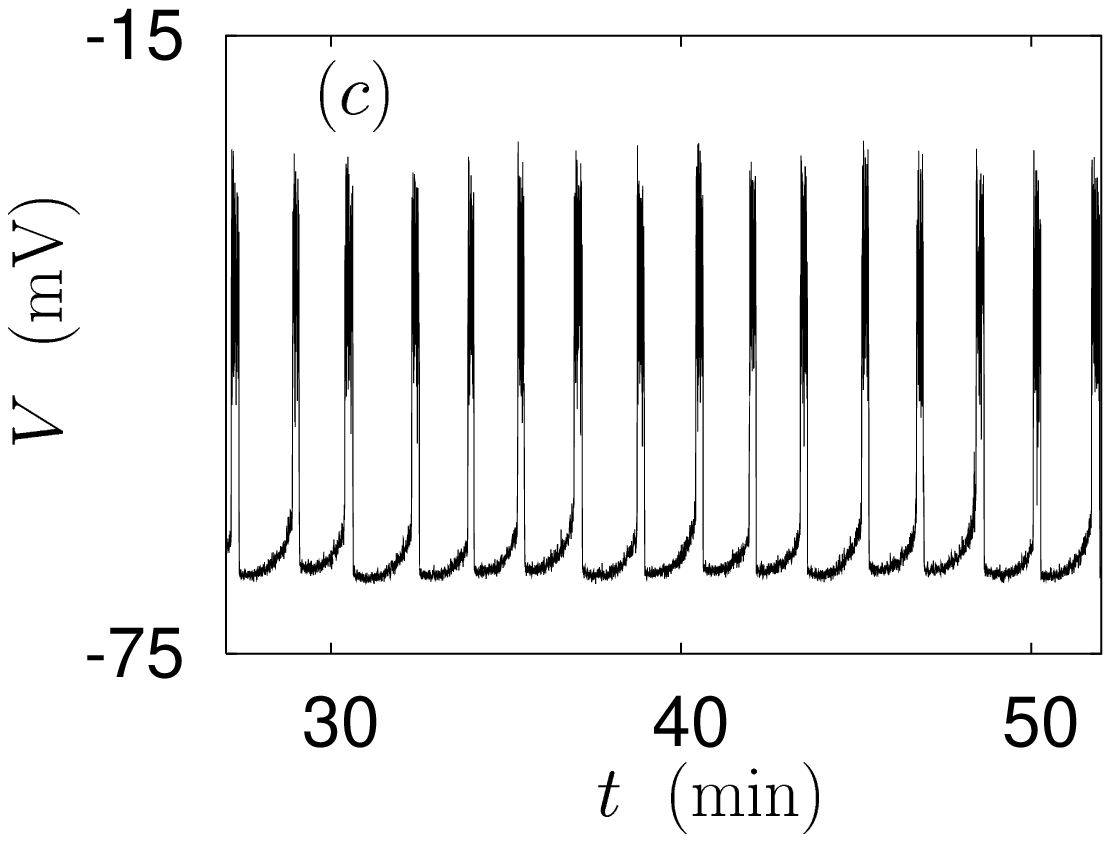} &
\epsfig{width=4.2cm,height=3cm,file=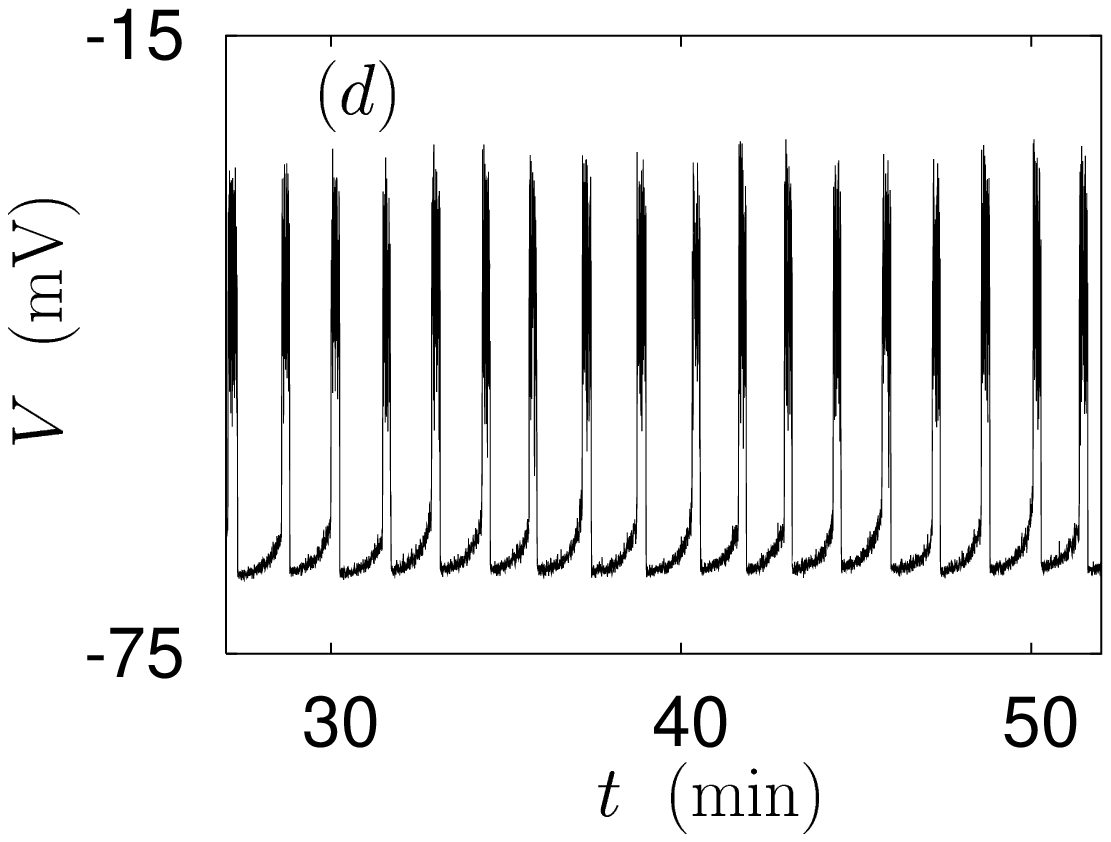} \\
\end{tabular}
\caption[membrane potential observed in direct effect to $g_{(K(ATP)}$
at several extracellular glucose concentrations]{Bursting behavior of the membrane
potential via the direct pathway mechanism with $\tau_J=3$\,min, at the same extracellular
glucose concentrations as in Fig. 3.
}
\label{Dir_V}
\end{figure}

\begin{figure}
\epsfig{width=7.16cm,height=5.7cm,file=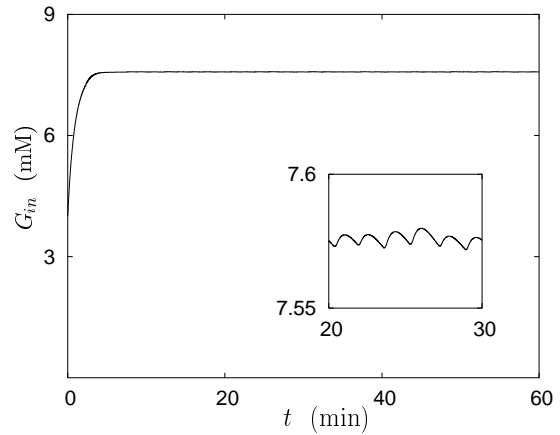} \\
\caption[intracellular glucose concentration at the extracellular glucose concentration
is $10$\,mM in direct effect condition]{Behavior of the intracellular glucose concentration
at $G_0=9$\,mM and $\tau_J=3$\,min in the presence of the direct pathway.
The inset shows an enlarged view.}
\label{Dir_Gi}
\end{figure}

\begin{figure}
\epsfig{width=7.16cm,height=5.7cm,file=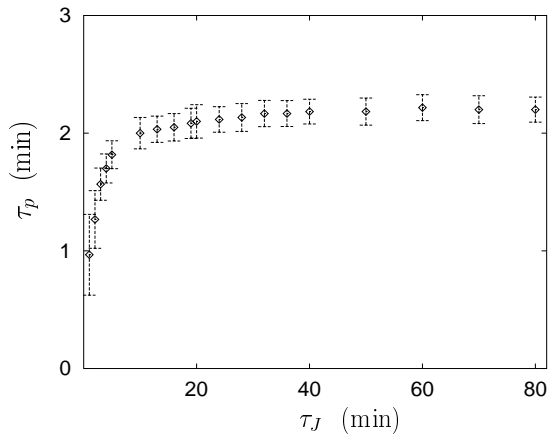} \\
\caption[period of bursting at various tauj]{
Bursting period $\tau_p$ depending on $\tau_J$ at $G_0=10$\,mM.}
\label{burstperio_tauj}
\end{figure}

There are reports that insulin itself inhibits insulin secretion in $\beta$-cells
although the exact mechanism is not clear~\cite{Persaud,Khan}.
Specifically, experimental data seem to suggest that the negative feedback of insulin acts
directly on ATP-dependent K$^+$ channels, suppressing action potential firing~\cite{Khan}.
To examine this direct pathway, we set $f_I=1$, $f_D=f$ with $J_0=0.15$, and
$H_J=1.0\times10^{-3}$\,mM, varying the relaxation time $\tau_J$.
Figure~\ref{Dir_V} exhibits the resulting oscillatory behaviors of the membrane potential
for $\tau_J = 3$\,min, depending on the extracellular glucose concentration $G_0$.
The average duration of bursting is rather insensitive to $G_0$,
approximately 16 seconds for a wide range of $G_0$.
Compared with the results of the indirect pathway, the duration time is short and
the dose-response to glucose is also weak in the direct pathway.
On the other hand, the average period between bursts decreases with $G_0$,
which results in more insulin secretion at higher glucose levels.
Unlike the indirect pathway, however, the clustering of bursts
does not emerge in the direct pathway. Since the secreted insulin affects
the ATP-dependent K$^+$ channels directly and rapidly, there is not enough time to activate
the mechanism for bursting clusters.
The behavior of the intracellular glucose concentration $G_{in}$ is shown in Fig.~\ref{Dir_Gi},
for the extracellular glucose concentration $G_0 = 9$\,mM and the relaxation time
$\tau_J = 3$\,min.
No prominent slow oscillations are observed (compare with Fig.~\ref{Gi_s}).
Figure~\ref{burstperio_tauj} shows how the average bursting period $\tau_p$ depends
on the relaxation time $\tau_J$: Observed is a rapid increase of $\tau_p$,
followed by saturation to $2$\,min, as $\tau_J$ is raised.

\begin{figure}
\epsfig{width=7.16cm,file=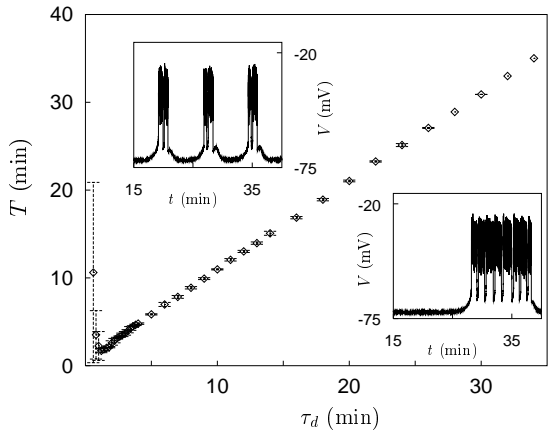}
\caption{Duration time $T$ of a bursting cluster versus the time delay $\tau_d$,
with typical error bars estimated from the standard deviation.
The insets exhibit the bursting patterns at $\tau_d = 1.2$\,min (upper left) and
$9.0$\,min (lower right).
}
\label{dubustatdir}
\end{figure}

To obtain bursting clusters via the direct pathway, one needs another formal time delay
and accordingly, an additional process in insulin inhibition.
As an attempt, we consider delay $\tau_d$ in the direct activation factor,
i.e., $f_D(t-\tau_d)$ on the conductance of the ATP-dependent K$^+$ channel,
and show the results in Fig.~\ref{dubustatdir}.
It is observed that the total duration time $T$ of a bursting cluster grows almost
linearly with the delay $\tau_d$.  When $\tau_d$ is small, we have $T\approx 0$ and
there is no regular bursting patterns.  When $\tau_d$ is raised to $1.2$\,min,
clustering patterns of bursts begin to appear; the corresponding behavior of the
membrane potential is displayed in the upper inset panel.
As $\tau_d$ is increased further, the bursting duration time keeps growing and
the bursting pattern at $\tau_d=9.0$\,min becomes similar to that via the
indirect pathway shown in Fig.~\ref{poten_s}(c).
Nevertheless it is rather difficult to recognize this direct pathway as an
appropriate mechanism for bursting clusters
since such large delay is apparently not consistent with experimental results~\cite{Khan}.

\section{Discussion}

Our model has employed the glucose-insulin feedback for slow dynamics, responsible for
slow oscillations in the system.
Unlike the fast oscillations of bursting action potentials, however,
the origin of the slow oscillations with periods about 5 to 10 minutes
is not understood well and still in controversy.
One of the main difficulties lies in identifying the primary oscillation
among many secondary ones.  Indeed in a pancreas, many chemical ingredients,
e.g., Ca$^{2+}$, O$_2$, ATP, glucose, and insulin,
show oscillatory behaviors with periods of several minutes~\cite{Liu,Chou,Porksen,metaosc}.
Recently, it has been reported that glycolysis produces spontaneous oscillations
in $\beta$-cells: Here Ca$^{2+}$ acts as a mediator,
transducing oscillatory metabolism into oscillatory secretion~\cite{metaosc}.
After completion of the first version of the manuscript, we became aware of the recent work,
which addresses the multiple bursting mode in pancreatic islets, with glycolysis
adopted as the slow dynamics~\cite{Bertram}.
In the present study, on the other hand, we have proposed another possible mechanism
for slow oscillations, namely, the negative feedback through the product (i.e., insulin)
in response to the stimulus (glucose).

In combining the slow dynamics (glucose-insulin feedback) with the fast dynamics
(bursting action potentials), we have allowed both dynamics to interact bidirectionally.
As a result, there emerge mixed oscillations of both time scales, clusters of bursts,
when the two dynamics complement each other via the indirect pathway.
This also keeps parallel with the glycolytic model~\cite{Bertram}, and contrasts with
the unidirectionally interacting model where slow dynamics of glycolysis influences
fast dynamics one-sidedly~\cite{Wierschem}.

We now discuss the plausibility of our model, in comparison with others.
Insulin secretion is synchronized among the $\beta$ cells in an islet, which has been observed
in the calcium imaging experiment~\cite{synchrony}.  In addition, the pulsatile insulin
secretion in a group of islets indicates additional synchronization among islets in a pancreas;
otherwise the secreted insulin from many islets, having different phases from each other,
would produce essentially a constant level of insulin~\cite{Pedersen},
although there may exist a source of some controversy~\cite{Valdeolmillos}.
Such synchronization among islets may not be explained by the model adopting glycolysis
as the slow dynamics, where no communication is available between islets.
An expanded model was also proposed, incorporating glucose release in the liver:
Secreted insulin causes oscillations of the plasma glucose through the liver,
which in turn entrain all islets with its characteristic frequency~\cite{Pedersen}.
This is, however, in disagreement with pulsatile insulin secretion at constant
glucose concentration observed in vitro~\cite{Stagner} and in vivo~\cite{Song}.
On the other hand, in our feedback model, secreted insulin from an islet affects other islets
(via the paracrine interaction) as well as the islet itself (via the autocrine interaction),
thus inherently allowing synchronization among islets even at a constant plasma glucose level.
Accordingly, our model can explain the sustained insulin oscillations at constant glucose
concentration observed in vitro and in vivo~\cite{Stagner,Song}.
Further, the recent observation that defects in IRS-1, 
associated with the reduced expression of the SERCA protein, 
result in lack of slow calcium oscillations and reduction of insulin secretion~\cite{Kulkarni} 
is also supportive of the insulin feedback mechanism. 
Note, however, that the molecules for negative feedback are not necessarily limited to insulin and
other products, e.g., GABA, can also provide the feedback pathway~\cite{Braun}.

In constructing a model for such complex biological systems, one may not expect perfection
and impeccableness, and there are certainly some drawbacks in our feedback model as well.
First, our model predicts that both the amplitude and the period of the insulin oscillation
increase with the plasma glucose concentration, which apparently does not agree with the report
that mainly the amplitude tends to change~\cite{Porksen,Song}.
Secondly, slow oscillations are still observed in experiment on single $\beta$ cells,
even when the flow rate of the local perfusion is too fast for secreted insulin to reach
the insulin receptor and to operate the feedback mechanism~\cite{Grapengiesser}.
In this case the other model adopting glycolysis as slow dynamics is still applicable
since the glycolytic oscillation may still occur without the feedback through a product like
insulin~\cite{Bertram}.
%
%

In summary, we have incorporated the macroscopic description of glucose regulation and
the microscopic mechanism of bursting behavior of $\beta$ cells, to establish a model
that displays inherently the observed oscillations of the membrane potential, cytosolic
calcium concentration, and insulin secretion in pancreatic islets.
In view of the experimental observations, the ATP-dependent conductance $g_{K(ATP)}$ of
the K$^+$ channel has been taken as a decreasing function of the glucose concentration
whereas the insulin secretion rate has been given by an increasing function of the
intracellular Ca$^{2+}$ concentration.
Considered here are two possible insulin induced inhibitory pathways, affecting
the conductance $g_{K(ATP)}$.
Further, $\beta$ cells in an islet have been considered to be coupled electrically
with their nearest neighbors with appropriate coupling conductance.
By means of extensive numerical simulations, we have obtained bursting electrical activities,
calcium concentration, and insulin secretion, which are consistent with those observed experimentally.
In addition to the stationary-state behaviors,
we have also examined how the intracellular glucose concentration changes
with the extracellular glucose concentration, and explored the behaviors of the system
as the injected glucose concentration is varied successively.
Then the inhibition effects of insulin on the bursting action potential have been probed.
%
Finally, we have compared two possible pathways to the feedback modulations of $g_{K(ATP)}$
by secreted insulin.
With appropriate parameters, the indirect feedback mechanism has been found to generate
bursting clusters of action potentials.
On the other hand, the direct mechanism in general generates only single bursts;
it requires, e.g., rather unrealistically long delay time to produce bursting clusters.

Our model may thus serve as a useful framework to predict insulin secretion
modulated by the molecular mechanism in $\beta$ cells
and the corresponding glucose regulation.
In particular, our model, based on the glucose-insulin feedback, can be a good complement
to the glycolytic model.
Extensions toward a more realistic description of the biological situation are
left for further study.

\section*{Acknowledgments}
We thank S.-W. Han for useful discussions at the early stage of this work.
This work was supported in part by the Korea Science and Engineering Foundation
through Grant No. R01-2002-000-00285-0 and through National Core Research Center
for Systems Bio-Dynamics.


\begin{thebibliography}{99}
\bibitem{index} 
For simplicity of notation, the subscript $i$ labeling the cell has been omitted.
\end{thebibliography}
\end{document}